\documentclass[useAMS,usenatbib]{mn2e}

\usepackage{epsf,rotating}
\usepackage{amsmath}
\usepackage{subfig}

\newcommand{\kmsmpc}{\kms\;{\rm Mpc}^{-1}}

\newcommand{\hkpc}{h^{-1}{\rm kpc}}
\newcommand{\hmpc}{h^{-1}{\rm Mpc}}

\newcommand{\kms}{\;{\rm km}\,{\rm s}^{-1}}

\newcommand{\gad}{{\sc Gadget-3}}
\newcommand{\gizmo}{{\sc Gizmo}}
\newcommand{\mufasa}{{\sc Mufasa}}

\newcommand{\ion}[2]{\hbox{#1\,{\sc #2}}}

\title[The {\sc Mufasa} Simulations]{{\mufasa}: Galaxy Formation Simulations With Meshless Hydrodynamics}

\author[Dav\'e, Thompson, \& Hopkins]{
\parbox[t]{\textwidth}{\vspace{-1cm}
Romeel Dav\'e$^{1,2,3}$, Robert Thompson$^{4,1}$, Philip F. Hopkins$^{5}$}
\\
\\$^1$ University of the Western Cape, Bellville, Cape Town 7535, South Africa
\\$^2$ South African Astronomical Observatories, Observatory, Cape Town 7925, South Africa
\\$^3$ African Institute for Mathematical Sciences, Muizenberg, Cape Town 7945, South Africa
\\$^4$ National Center for Supercomputing Applications, Champaign-Urbana, IL 61801
\\$^5$ California Institute of Technology, Pasadena, CA 91125
}

\begin{document}

\maketitle

 \begin{abstract}
We present the \mufasa\ suite of cosmological hydrodynamic simulations,
which employs the {\sc Gizmo} meshless finite mass (MFM) code
including $H_2$-based star formation, nine-element chemical evolution,
two-phase kinetic outflows following scalings from the Feedback in
Realistic Environments zoom simulations, and evolving halo mass-based
quenching.  Our fiducial $(50\hmpc)^3$ volume is evolved to $z=0$
with a quarter billion elements. The predicted galaxy stellar mass
functions (GSMF) reproduces observations from $z=4\rightarrow 0$
to $\la 1.2\sigma$ in cosmic variance, providing an unprecedented
match to this key diagnostic.  The cosmic star formation history
and stellar mass growth show general agreement with data, with a
strong archaeological downsizing trend such that dwarf galaxies
form the majority of their stars after $z\sim 1$.  We run $25\hmpc$
and $12.5\hmpc$ volumes to $z=2$ with identical feedback prescriptions,
the latter resolving all hydrogen-cooling halos, and the three runs
display fair resolution convergence.  The specific star formation
rates broadly agree with data at $z=0$, but are underpredicted at
$z\sim 2$ by a factor of three, re-emphasizing a longstanding puzzle
in galaxy evolution models.  We compare runs using MFM and two
flavours of Smoothed Particle Hydrodynamics, and show that the GSMF
is sensitive to hydrodynamics methodology at the $\sim\times 2$
level, which is sub-dominant to choices for parameterising feedback.
\end{abstract}

\begin{keywords}
galaxies: formation, galaxies: evolution, methods: N-body simulations, galaxies: mass function
\end{keywords}

\section{Introduction}

Galaxies were first identified over a century ago, yet the physical
processes that shape their observable characteristics and evolution
remain poorly understood.  With the growth of large-scale structure
now well specified within the concordance Lambda Cold Dark Matter
($\Lambda$CDM) paradigm, the least understood issues surround how
the baryonic components behave once the dark matter's gravity is
no longer the dominant driver.  In particular, much of the uncertainty
lies in the nature and physics of {\it feedback} processes, i.e.
the energy return from processes of star and black hole formation
within galaxies that self-regulates their growth.  As observations
of the galaxy population from the present day back to the epoch of
reionisation continue to improve, it is an increasingly demanding
challenge for models to reproduce such observations using well-motivated
and comprehensive input physics, while giving insights into the
physical processes responsible for feedback and galaxy assembly.

Successful modern galaxy formation models generally rely on the
following basic ingredients~\citep{Somerville-15}: (i) Fueling of
star formation via gravitationally-driven gas accretion from the
intergalactic medium (IGM); (ii) feedback from massive, young stars
that drives large-scale galactic outflows; and (iii) feedback from
supermassive black hole accretion that suppresses star formation
in massive galaxies.  The goal of galaxy formation models is thus
to understand and constrain the physics driving these various
feedback processes, and particularly how they work in concert to
shape galaxies as we observe them via multi-wavelength surveys
across cosmic time.

Cosmological hydrodynamic simulations have played a major role in
the progress towards understanding galaxy formation.  The past
decade has seen significant advances in both the size and dynamic
range of simulations thanks to advancing computing power, as well
as the scope and accuracy of the input physics thanks to an improved
understanding of the key physical drivers.  Such simulations are
now able to broadly reproduce primary demographic features of the
galaxy population today, along with their evolution to early epochs,
to reasonable levels of success
\citep[e.g.][]{Dave-11a,Dave-11b,Vogelsberger-14,Genel-14,Schaye-15,Somerville-15}.
However, all cosmological-volume models still require ``sub-resolution"
recipes for feedback processes, because despite heroic computational
efforts it remains infeasible to simultaneously resolve the physics
driving feedback at sub-parsec scales within a megaparsec-scale
structure formation context.  While current models are typically
tuned to reproduce key observations such as a galaxy stellar mass
function to within a factor of several, they accomplish this using
widely differing input physics and numerical implementations, making
it difficult to draw robust conclusions regarding the detailed
physical processes driving feedback.

One way forward in order to hone in on viable sub-resolution recipes
for feedback is to utilise the results of very high-resolution
simulations of individual galaxies that examine specific feedback
processes in detail.  This has recently been greatly enabled by the ``zoom"
simulation technique in which a particular galaxy and its surroundings
are extracted from a larger volume and re-simulated at much higher
resolution.  In this way, feedback processes are generated more
self-consistently and tracked more
directly~\citep[e.g.][]{Hopkins-14,Christensen-16}.  It is not
straightforward to predict population statistics from individual
zoom simulations, but their predictions can serve as inputs for the
recipes used in cosmological-scale simulations that model many
thousands of galaxies.

A particularly comprehensive set of physical processes was included
in the Feedback In Realistic Environments (FIRE) zoom
simulations~\citep{Hopkins-14}.  Using the FIRE simulations,
\citet{Muratov-15} derived scaling relations for the mass loss rate
and velocity distribution of outflowing gas, which are key aspects
of star formation-driven feedback.  Detailed models for
self-shielding~\citep[e.g.][]{Faucher-10,Rahmati-13}, black hole
growth~\citep[e.g.][]{Sijacki-07,Angles-15,Angles-16}, and
quenching~\citep{Gabor-11} have been investigated to study their
emergent trends with a similar strategy in mind.  These trends can
then be implemented into large-scale, lower-resolution simulations
to provide more physically motivated recipes for feedback.
Such a multi-scale approach offers a promising avenue to assemble
robust and well-constrained recipes for feedback processes within
cosmological-scale galaxy formation models.

Another important area of progress has been towards increasing the
robustness of hydrodynamics solvers.  Early simulations were
predominantly done using the Smoothed Particle Hydrodynamics (SPH)
technique~\citep{Hernquist-89,Monaghan-92}.  However, this method
requires substantial artificial viscosity that compromises the
handling of surface instabilities, contact discontinuities, and
strong shocks.  A reformulation of the SPH equations to explicitly
conserve entropy provides some advantages~\citep{Springel-02,Springel-05},
but still resulted in numerical surface tension that e.g. strongly
supresses Kelvin-Hemholtz instabilities~\citep{Agertz-07}.  Further
reformulations such as SPH-S~\citep{Read-12} and
Pressure-SPH~\citep{Hopkins-13} have partially but not fully mitigated
these issues.

Adaptive mesh refinement (AMR) galaxy formation codes have also
made substantial gains~\citep[e.g.][]{Enzo-14}, but their intrinsically
Cartesian nature can result in numerical artifacts when applied to
the arbitrary 3-D geometry and motions associated with galaxy
formation.  The Voronoi tesselation-based moving mesh code {\sc
Arepo}~\citep{Springel-10} eliminates the Cartesian restrictions,
and shows great promise at handling surface instabilities and shocks
with arbitrary geometries while retaining Galilean invariance.  A
downside is that {\sc Arepo} advects mass across cell boundaries
and occasionally must ``re-mesh" to prevent highly deformed cells.
This necessitates the additional complication of tracer particles
in order to track the motion of gas into and out of
galaxies~\citep{Genel-14}.

Recently, \citet{Hopkins-15a} developed a new ``meshless" hydrodynamics
method that marries many advantages of SPH and moving mesh approaches.
{\sc Gizmo} is fundamentally a moving mesh-like Godunov code, but
the ``mesh" is defined by a deformable kernel moving in such a way
as to keep the mass within each cell constant.  Hence operationally
it allows for the advantages of mass tracking and being fully
adaptive without re-meshing.  \citet{Hopkins-15a} showed that {\sc
Gizmo's} ``meshless finite mass" (MFM) approach shows desirable
behavior compared to modern SPH and AMR codes in many relevant
idealised test problems\footnote{\tt
http://www.tapir.caltech.edu/\~{}phopkins/Site/GIZMO.html}.

In this paper we present a new suite of cosmological hydrodynamic
simulations of galaxy formation using {\sc Gizmo}, called the
\mufasa\ simulations.  We describe our new recipes for sub-resolution
processes (\S\ref{sec:code}) based on, wherever possible, the results
from zoom simulations or analytic models.  These include $H_2$-based
star formation, minimal ISM pressurization, non-equilibrium cooling
and ionisation, supernova-heated two-phase kinetic winds, evolving
halo mass-based quenching, and nine-metal chemical enrichment from
Type Ia+II supernovae and stellar evolution.  Our highest resolution
\mufasa\ run resolves essentially all hydrogen cooling halos, while
our fiducial $50\hmpc$ volume produces a substantial population of
quenched massive galaxies by $z=0$.  All are run with exactly the
same feedback parameters, and show reasonable resolution convergence.
As a basic test of our model, we present predictions for the galaxy
stellar mass function across cosmic time, and compare to recent
observations and simulations (\S\ref{sec:gsmf}).  We further compare
to observations of the star formation histories, both globally and
for individual galaxies (\S\ref{sec:sfr}).  We also examine the how
the choice of hydrodynamics solver and key feedback parameters
impact our results (\S\ref{sec:codevar}).  Finally, we summarize
our findings (\S\ref{sec:summary}).

\section{Code Description}\label{sec:code}

\subsection{{\sc Gizmo} gravity and hydrodynamics}

We employ a modified version of the gravity plus hydrodynamics
solver {\sc Gizmo}~\citep{Hopkins-15a}, which is built upon the
framework of the \gad\ galaxy formation code~\citep{Springel-05}.
For hydrodynamics, we employ {\sc Gizmo}'s meshless finite mass
(MFM) method, which uses a Riemann solver to evolve the fluid in a
manner that conserves mass within each fluid element.  Each fluid
element may then be regarded as a ``particle", though it actually
represent a mesh node; we will refer to these as ``gas elements."
Like {\sc Arepo}, {\sc Gizmo} offers advantages over SPH in terms
of handling a wide range of fluid problems related to shocks and
contact discontinuities, and over AMR codes in terms of avoiding
Cartesian artifacts and strictly preserving Galilean invariance.

We employ a cubic spline kernel with 64 neighbors in MFM.
\citet{Hopkins-15b} used 32 neighbors for their detailed MHD
simulations, but we discovered that 64 is better for suppressing
occasional instabilities in very diffuse hot gas such as that
appearing in the shocked IGM at late epochs.  The kernel is used
to determine the volume partition between gas elements, thereby defining
the effective faces at which the Riemann problem is solved.

In order to test the sensitivity to hydrodynamics methodology, we
will also use the modern Pressure-SPH~\citep[P-SPH;][]{Hopkins-13}
and the Traditional (entropy-conserving) SPH~\citep[T-SPH;][]{Springel-02}
solvers in {\sc Gizmo}, with all subgrid physics options held fixed.
SPH requires an artificial viscosity, for which we use the {\sc
Gizmo} default \citet{Cullen-10} viscosity as described in
\citet{Hopkins-15a}.  To isolate the impact of the hydro solver,
we use the same cubic spline kernel with 64 neighbors, even though
a quintic spline gives better results in some circumstances
\footnote{\citet{Dehnen-12} found that SPH similar to T-SPH using
a cubic spline kernel with 64 neighbors may, in strong shearing
flows, be subject to particle pairing instabilities that result in
lowering the effective resolution.  We thus ran the same T-SPH
simulation with a quintic spline with 64 neighbors, and the results
were essentially identical.  Hence this pairing instability is
evidently not impacting our results.}.

The gravity is evolved using a Tree-Particle-Mesh approach that is
based on \gad~\citep{Springel-05}.  We use adaptive gravitational
softening throughout for all particles~\citep{Hopkins-15a}, enclosing
64 particles with a minimum (Plummer-equivalent) softening length
set to 0.5\% of the mean interparticle spacing.  For more details
on {\sc Gizmo}, including other choices such as the timestep
limiter~\citep{Durier-12} and the implementation of adaptive
softening, see \citet{Hopkins-15a}.

\subsection{Radiative Cooling and Heating}

We use the {\sc Grackle-2.1} chemistry and cooling
library~\citep{Enzo-14,Kim-14}, which includes primordial and metal
line cooling, evolved isochorically on the cooling timescale in an
operator-split way over the system timestep.  In other words, we
do the non-radiative cooling/heating first over the system timestep,
then apply radiative cooling using {\sc Grackle}, where the radiative
part is sub-stepped isochorically on the cooling timescale.

We employ {\sc Grackle} in non-equilibrium mode for primordial
elements, thereby tracking e.g. the input of latent heat during
reionisation.  Metal-line cooling is computed using {\sc
Cloudy}~\citep{Ferland-04} tables self-consistently including the
impact of photoionisation, in this case assuming ionisation
equilibrium.  A spatially-uniform photo-ionising background is
assumed, for which we employ the determination by \citet{Faucher-09},
updated in 2011.  We do not employ the more recent \citet{Haardt-12}
determination owing to concerns regarding its low \ion{H}{i}
ionisation rate at $z=0$ which has spurred the so-called photon
underproduction crisis~\citep{Kollmeier-14}.  In any case, galaxy
properties that we consider here are insensitive to reasonable
choices of the photo-ionising background.

\subsection{Star Formation}

To form star particles out of gas, we employ a molecular gas-based
star formation prescription following \citet[hereafter KMT]{Krumholz-09}.
A fuller description is available in \citet{Krumholz-11} and
\citet{Thompson-14}; here we briefly summarise the approach.

KMT provides an approximate solver for $H_2$ formation at the
(relatively) coarse resolution available to us in cosmological
volumes.  
With well-motivated approximations, this yields
\begin{equation}
f_{\rm H2} = 1 - 0.75{s\over 1+0.25s}
\end{equation}
where 
\begin{equation}
s = {\ln (1+0.6\chi+0.01\chi^2)\over 0.0396 Z (\Sigma/M_\odot {\rm pc}^{-2})},
\end{equation}
where $Z$ is the metallicity in solar units, $\chi$ is a function
of metallicity given in KMT, and $\Sigma$ is the
column density calculated using the Sobolev approximation; see below.
We impose a minimum metallicity of $10^{-3}Z_\odot$ solely for the
purposes of this computation.  We further require a minimum
density for star formation of $n_H\geq 0.13$~cm$^{-3}$; gas above
this density will hereafter be referred to as ``ISM gas".  In most
cases, the densities at which $H_2$ dominates is higher
than this (e.g. $n_H\sim 0.5-1$~cm$^{-3}$ for solar
metallicity) but for individual high metallicity gas elements the KMT
formula can yield $H_2$ formation, and hence star formation, below our 
ISM density threshold; we explicitly do not allow this.

We make two minor adjustments to the KMT algorithm regarding sub-grid
clumping.  First, KMT
assumed a constant sub-resolution clumping factor in the ISM of
${\cal C}=30$.  This factor appears in the rate coefficient for
forming $H_2$ on dust grains~\citep[see \S2.2 of][]{Krumholz-11}.
It is clear that a higher resolution simulation should assume
less clumping -- in the limiting case of infinite resolution, the
clumping should be fully resolved and hence ${\cal C}$ should be
unity.  Thus it makes physical sense to reduce ${\cal C}$ in higher
resolution simulations.  In our case, we assume that ${\cal C}=30$
in our fiducial simulation with $\epsilon=0.5\hkpc$ (see
\S\ref{sec:runs}), where $\epsilon$ is our minimum comoving
gravitational softening length, but we further scale ${\cal C}\propto
\epsilon$ in our other runs (see Table~\ref{tab:sims}).  By this
scaling, a simulation with $\epsilon\sim 25$~pc would have ${\cal
C}=1$; our cosmological runs here do not reach this limit.  The net
effect of scaling ${\cal C}$ is to make star formation slightly
less efficient in higher resolution simulations, leaving
them slightly more gas rich.

Our second adjustment involves the calculation of $\Sigma$.  KMT
assumes that self-shielding against $H_2$ dissociating radiation
within the ISM is governed by the local column density $\Sigma$ as
calculated using the Sobolev approximation, namely
$\Sigma=\rho^2/|\nabla\rho|$ where $\rho$ is the gas density.  We
compute these quantities in the usual way from the gas
elements, but we must further account for sub-resolution clumping.
If the clumping factor ${\cal C}$ represents the typical density
contrast of the sub-resolution clumps, then the column density
increase within these clumps is approximately, by dimensional
arguments, ${\cal C}^{2/3}$.  We thus increase $\Sigma$ by a factor
of ${\cal C}^{2/3}\approx 10$ to account for this\footnote{ The
factor of 10 is appropriate for our fiducial large volume.  This
factor should vary as ${\cal C}$ varies, but owing to an oversight
this was not implemented for the higher-resolution runs.  We
subsequently ran test simulations to determine that the results are
not significantly different if we scale $\Sigma\propto {\cal
C}^{2/3}$, though it does lower star formation very slightly.  Future
runs will use this scaling.}.

Given an $H_2$ fraction from KMT, the star formation rate for a
given eligible gas element is given by a \citet{Schmidt-59} Law,
namely
\begin{equation}
\frac{dM_*}{dt} = \epsilon_* \frac{\rho f_{\rm H2}}{t_{\rm dyn}},
\end{equation}
where $f_{\rm H2}$ is the molecular fraction for a given gas element,
and $t_{\rm dyn}=1/\sqrt{G\rho}$ is the local dynamical time.  We set
the efficiency of star formation to be $\epsilon_*=0.02$~\citep{Kennicutt-98}.

Star particles are spawned from gas elements stochastically.  Each
MFM fluid element spawns a single star particle of the same mass.
Previous simulations~\citep[e.g.][]{Oppenheimer-08} often spawn
multiple star particles, reducing the gas mass by a fraction each
time.  However, large instantaneous changes to an element's gas
mass can cause instabilities in MFM, so we avoid this by converting
the gas element fully into a collisionless star particle all at
once, conserving mass and momentum.

\subsection{Resolving the Jeans mass in the ISM}

As the density increases and the cooling rate becomes large,
the Jeans mass can become smaller than the resolved mass in the
simulation.  If cooling is allowed to proceed at these high densities,
the fragmentation that should proceed to small scales (and thus be
captured by our star formation law) is instead ``bottlenecked" at
large scales, producing artificially coherent clumping~\citep{Schaye-08}.

To avoid this, we apply an artificial pressure above a density
threshold $n_{\rm th}$ that explicitly suppresses fragmentation below
the scale of the smoothing volume, namely:
\begin{equation}
n_{\rm th} = \frac{3}{4\pi\mu m_p} \Bigl(\frac{5k_BT_0}{G\mu m_p}\Bigr)^3 \Bigl(\frac{1}{N_{\rm ngb}m_g}\Bigr)^2,
\end{equation}
where $\mu$ is the molecular weight calculated assuming fully neutral gas, 
$m_g$ is the mass per gas element in our simulation, and $N_{\rm ngb}=64$
is the number of neighbors.

We assume that the temperature at the density threshold is $T_0=10^4$K.
Above this density, we then set the gas element temperature to the 
{\it Jeans minimum temperature:}
\begin{equation}
T_{\rm JMT}=T_0 (n/n_{\rm th})^{1/3}.
\end{equation}
If $n_{\rm th}$ is below our nominal star formation density
threshold, we set it to this value; this occurs in our largest volume.
Meanwhile, our smallest volume has $n_{\rm th}=11.6$~cm$^{-3}$, while our intermediate
volume has $n_{\rm th}=0.18$~cm$^{-3}$.

This prescription is quite similar to the $P\propto T^{4/3}$ equation
of state imposed by \citet{Schaye-08}, except that we vary the
threshold density with the simulation resolution such that at higher
resolution, we do not artifically pressurise the ISM as strongly.
Hence our approach automatically takes advantage of high numerical
resolution, when present, as much as possible~\citep[e.g.][]{Thompson-16}.

\subsection{Feedback From Massive Stars}

As necessitated by our cosmological resolution, we employ a
sub-resolution approach to modeling feedback from massive stars.
The basic approach is to use kinetic
feedback~\citep{Springel-03,Oppenheimer-06}, as this has generally
been shown to provide stable and well-converged results for
cosmological galaxy formation simulations, and allows more explicit
control over the numerical experiments regarding feedback.

In our feedback prescription, we assume that massive stars launch
winds that drive material out of galaxies through a combination of
Type~II supernovae (SNII), radiation pressure, and stellar winds.
We do not track these phenomena in the code owing to a lack of
resolution, but rather characterise the net effect using two free
parameters, namely the mass loading factor $\eta$ defined as the
mass outflow rate relative to the star formation rate, and the wind
speed $v_w$ representing the velocity with which the wind is launched.
Both parameters can in principle scale with galaxy properties and
redshift.  Outflow fluid elements are launched in the direction of
$\pm${\bf v}$\times${\bf a}, in order to have a broadly collimated
outflow perpendicular to the disk, and the sign is chosen such that
it is launched in the hemisphere pointing away from the host galaxy.
Later we will show that this choice has no discernible impact versus
ejecting gas isotropically.

\subsubsection{FIRE wind scalings}

Our general strategy is to directly employ outflow scalings derived
from the FIRE simulations~\citep{Hopkins-14}, which self-consistently
drive winds from galaxies using supernova feedback, radiation from
massive stars, and stellar winds, and have been successful at
reproducing a range of observed galaxy properties in individual
galaxy zoom simulations.

\citet{Muratov-15} found in the FIRE simulations that the outflow
rate was most tightly correlated with stellar mass, and this scaling
was essentially independent of redshift.  Their best-fit relation
of the mass outflow rate at $0.25R_{\rm vir}$ is given by
\begin{equation} \label{eq:eta}
\eta=3.55 \Bigl(\frac{M_*}{10^{10}M_\odot}\Bigr)^{-0.351},
\end{equation}
where $M_*$ is the galaxy stellar mass.  We implement this relation
directly into our code.  We compute $M_*$ using an on-the-fly galaxy finder, which 
we describe in \S\ref{sec:fofrad}.  To account for the very early evolution
of galaxies when little or no stellar mass has yet formed, we place
a floor on the stellar mass used in the above equation of 8~gas
element masses.

For the wind speed, we likewise take the scaling of $v_w$ with
galaxy properties from the FIRE simulations.  Their best-fit formula
is $v_w = 0.854 v_c^{1.12}$, where $v_c$ is the galaxy's circular
velocity.  We will describe how we obtain $v_c$ from our on-the-fly
galaxy finder in \S\ref{sec:fofrad}.  In FIRE, this formula reflects
the mean velocities across a shell at $0.25R_{\rm vir}$.  To account
for the fact that our wind launches are typically at a much smaller
radius well within the ISM, we add an extra velocity kick $\Delta
v_{0.25}$ corresponding to the gravitational potential difference
between the wind launch radius and $0.25R_{\rm vir}$~\citep{Lokas-01}.
We assume a \citet{Navarro-97} profile with a concentration 
$c=9(M_{\rm halo}/10^{10} M_\odot)^{-0.15}$, where we compute the
halo mass as the galaxy baryonic mass times $\Omega/\Omega_b$.
This typically increases the wind velocity by up to $\sim v_c$ in small
galaxies, and by significantly less in large galaxies.

Still, $\Delta v_{0.25}$  only accounts for gravitational slowing,
while hydrodynamic slowing is likely to be dominant~\citep{Oppenheimer-08}.
Hence we further increase the multiplier of the outflow velocity,
by around a factor of two above the \citet{Muratov-15}
value, comparable to their 95th percentile wind velocity, and also
comparable to the outflow velocities measured leaving the disk in
the zoom simulations of \citet{Christensen-16}.  We choose to
normalize this multiplier to the FIRE scaling at $200 \kms$.  Hence the
velocity formula used in \mufasa\ is 
\begin{equation}\label{eq:vw}
v_w = 2 \Bigl({v_c\over 200}\Bigr)^{0.12} v_c + \Delta v_{0.25}.
\end{equation}
Finally, we add a random tophat scatter of $\pm 10\%$ to $v_w$.
The resulting wind launch velocities are typically $\sim 2-3 v_c$,
which is comparable to the observed maximum outflow speeds in
low-redshift~\citep{Martin-05} and high-redshift~\citep{Steidel-10}
galaxies.  

\subsubsection{Two-phase winds}

Winds are often observed to be flowing out of nearby galaxies (e.g.
M82) partly in hot gas, putatively heated by supernova explosions
and the resulting shocks.  Previous kinetic wind
models~\citep{Dave-11a,Vogelsberger-14} have generally not included
this form of energy injection, even though their simulations lack
the resolution to properly capture SNe shock heating in the ISM.
Winds are also seen in cool form, mainly via low-ionisation absorption
lines.  Hence it is clear that there are at least two phases for
outflowing gas.

In \mufasa, we model two-phase winds by randomly selecting outflowing gas
to be ``cool" or ``hot".  Cool gas is ejected at $T\approx
10^3$~K, although they quickly heat to $\sim 10^4$~K via photo-ionisation
once they are outside the ISM.  For the hot wind, we directly employ a fraction
of the supernova energy not used for kinetic ejection into thermal
heating of the wind.  For a Chabrier initial mass function (IMF),
each solar mass of stars produces 0.01028 stars with masses above
$8M_\odot$.  We assume each of these becomes a supernova which
produces $10^{51}$~ergs, hence the specific energy production from
SNe is $u_{SN}=5.165\times 10^{15}$~erg~g$^{-1}$.

The kinetic energy used by the wind is $\frac{1}{2}m_wv_w^2 =
\frac{1}{2} \eta m_g v_w^2$, where we assume that the total wind
mass ejected owing to a gas mass of $m_g$ forming into stars is
$\eta m_g$.  If each wind fluid element has mass $m_g$, then the specific
energy utilised by the wind is $\frac{1}{2}\eta v_w^2$.

The specific thermal energy added to each hot wind gas element is then
\begin{equation}
\Delta u = u_{SN}-\frac{1}{2}\eta v_w^2.
\end{equation}
If $\Delta u<0$ we add no heat; physically, this can represent a
situation where other wind mechanisms such as radiation pressure
are primarily responsible for driving winds.  We have found that
so long as the heating raises the typical temperature above a few
times $10^5$K to get it past the Helium and \ion{O}{vi} cooling peaks
(which $\Delta u$ almost always does), then the impact on galaxies
is relatively insensitive to the exact amount of heating.

We define a free parameter $f_{\rm hot}$ that determines the fraction
of the wind that is ejected hot.  Each wind element is then randomly
chosen to be either hot or cool, based on the probability $f_{\rm
hot}$.  In \mufasa, we choose $f_{\rm hot}=0.3$, motivated by
ISM-scale simulations of supernova-driven winds that suggest roughly
30\% of the material is ejected in hot form (T. Naab, priv. comm.).
Hence our two-phase wind is assumed to be majority cool, but
containing a substantial component of hot gas.

\subsubsection{Hydrodynamic decoupling and recoupling}

No hydrodynamics code is able to properly evolve a single fluid
element (cell or particle) moving with a high Mach number ${\cal
M}$ through its surroundings.  This is simply a limitation of
resolution, and is not mitigated by the use of an improved hydrodynamics
solver in MFM: A single fluid element will never be able to capture
the sub-element mixing and dissipation with its surrounding gas.
For this reason, it is necessary to make some approximation regarding
the interactions of the outflowing gas when passing through the
ISM.  Ideally, one would like to follow the detailed evolution of
gas mixing with its surroundings via surface instabilities, but
this is non-trivial, since the Kelvin-Hemholtz mixing time is
generally very short.  Instead, we follow early works by
\citet{Springel-03} and \citet{Oppenheimer-06,Oppenheimer-08}, along
with recent simulations such as Illustris~\citep{Vogelsberger-14},
and employ {\it decoupled} winds.  That is, we turn off hydrodynamic
forces on outflowing gas until such time as it ``recouples" back
to its surrounding gas.  The physical motivation is that winds are
often seen to blow channels through the ISM that likely allow the
relatively unfettered escape of a coherent flow.  Moreover, this
makes physical sense since our outflow scalings are taken from FIRE
simulations who measured their wind properties well outside the
ISM.

The criteria for recoupling has historically been chosen via a
density threshold, e.g. 10\% of the ISM density threshold, and a
time limit, e.g. enough to free-stream for tens of
kpc~\citep[e.g.][]{Springel-03}.  Here, we add a new recoupling
criterion: That the velocity of the outflowing gas element be similar to its
surroundings.  This criterion additionally allows recoupling when
the gas has a velocity that can be properly handled by the
hydrodynamics solver.

Specifically, our recoupling criterion requires that the wind
gas element has a velocity difference with respect to surrounding gas
of less than 50\% of the local sound speed (${\cal M}<0.5$).  The
sound speed is calculated from the kernel-weighted temperature of
surrounding gas at the wind element's position, and the velocity
difference is summed over all neighbors with kernel weighting.  We
exclude all other decoupled wind gas when computing quantities for the
surrounding gas.  Using 20\% or 100\% rather than 50\% did not yield
appreciably different results.  We further turn off radiative cooling
while the wind is decoupled, in order to allow hot gas in the wind
to deposit their thermal energy into the circum-galactic medium (CGM).

Gas elements satisfying the mach number criterion are recoupled
regardless of surrounding density or time, so long as they have a
density that is less than the ISM density.  But there are cases
where this recoupling criterion is not satisfied for a long while
or until far away, particularly in low-mass halos where the outflows
are typically ejected into very low density gas.  Hence we also
employ a density limit of 1\% of the ISM threshold density, and a
time limit of 2\% of the Hubble time at launch which is comparable
to a disk dynamical time; reaching either of these criteria
automatically triggers recoupling as well.  In the end, the total
fraction of decoupled gas is always well below 1\%,
reaching a maximum of around $0.2-0.3$\% at $z\sim 2$.

\subsection{Fast, Approximate Friends-of-Friends}\label{sec:fofrad}

Equation~\ref{eq:eta} requires $M_*$ to determine $\eta$, and
equation~\ref{eq:vw} requires $v_c$ to determine the outflow velocity.
To determine $M_*$ and $v_c$, we employ a fast approximate
friends-of-friend (FOF) galaxy finder on the fly during the simulation
run.  This FOF finder groups stars and ISM gas into baryonic galaxies.
It is approximate because it finds neighbours within a cube of side
$2{\cal L}$, where ${\cal L}$ is the linking length, rather than a
sphere of radius ${\cal L}$.  This greatly improves computation
speed while having minimal impact on the identified galaxies, thus
enabling the code to run this FOF finder at essentially every
timestep.  We calibrate ${\cal L}=0.0056$ times the mean interparticle
spacing in order to obtain the same mass function as that obtained
by using the Spline Kernel Interpolative Denmax~\citep[SKID;
e.g.][]{Keres-05} galaxy finder.  Our results are insensitive to
variations in ${\cal L}$ by a factor of a few around our chosen
value, but we choose this value since it is the minimum hydrodynamic
kernel size, which is 40\% of the minimum softening length, multiplied
by 2.8 which is the conversion from Plummer equivalent softening
to the actual smoothing kernel radius.  Hence we link particles
that are within one (minimum) kernel length of each other.

This FOF finder yields groupings of star particles and gas elements, from which we compute
the baryonic mass $M_b$ and $M_*$.  Given $M_b$, we then determine
$v_c$ for use in our outflow scalings by employing the observed
baryonic Tully-Fisher relation~\citep{McGaugh-12} with a redshift
evolution as expected from \citet{Mo-98}, namely
\begin{equation}
v_c = (M_b/102.329 M_\odot)^{0.26178} (H(z)/H_0)^{1/3},
\end{equation}
where $H(z)$ is Hubble's constant at redshift $z$.  



\subsection{Feedback From Long-Lived Stars}

Type Ia supernovae (SNIa) and asymptotic giant branch (AGB) stars
provide additional energetic feedback.  These components provide
feedback that is substantially delayed relative to the time of star
formation, hence are not represented by our kinetic outflows and
must be included separately.  Typically, delayed feedback is
energetically sub-dominant relative to the processes that drive
winds, but they may be important in particular circumstances, hence
we include their contributions for completeness.

The SNIa rate is modeled following \citet{Scannapieco-05} as a
prompt component that is concurrent with SNII, and a delayed component
that emerges from stars and begins after an age of 0.7~Gyr.  The
rates are taken from \citet{Sullivan-06}, namely $3.9\times 10^{-4}$
SNIa~$(M_\odot/$yr$)^{-1}$ for the prompt component, and $5.3\times
10^{-14}$ SNIa~$M_\odot^{-1}$ for the delayed component~\citep[see][for
further details]{Oppenheimer-08}.

Each SNIa is assumed to add $10^{51}$~erg of energy to the surrounding
gas.  The prompt component energy is added at each timestep to each
star-forming gas element, and the delayed component is added to
the 16 nearest gas elements to the given star, in a kernel-weighted
fashion.  To save computation time, we do the delayed feedback for
stars only at specific intervals, with the interval scaling roughly
inversely with the age of the star, from a minimum of every timestep
for ages under 200~Myr to a maximum of every 10 timesteps for stars
2~Gyr and older.

AGB star feedback also adds energy to surrounding gas.  We implement
the model of \citet{Conroy-15} in which AGB stellar winds are assumed
to thermalise with the ambient gas.  First, we compute the stellar
mass loss rate assuming a \citet{Chabrier-03} initial mass function
using a lookup table based on \citet{Bruzual-03} models.  We then
calculate the energy deposition rate based on the mass loss rate
and the velocity difference of the star relative to each neighboring
gas element.  We again use the 16 nearest neighbors, in a
kernel-weighted fashion.  We add the velocity of the stellar wind
to this, which we assume to be 100~km/s~\citep{Conroy-15}.  In
practice, even this large wind stellar velocity that is on the upper
end of plausibility makes a negligible difference for most galaxies.
This is because if the energy is added to star-forming gas it will
quickly radiate away.  However, such energy could potentially
accumulate into making a substantial impact if the stars are embedded
primarily in hot gas, such as in a cluster environment; we will
examine this in more detail in future work.

\subsection{Chemical Enrichment}

Delayed feedback mechanisms also deposit heavy elements into
the surrounding gas.  We track the evolution of 11 elements, including
hydrogen, helium, and 9 metals:  C, N, O, Ne, Mg, Si, S, Ca, and
Fe.  These elements comprise over 90\% of metal mass in
the universe.

For SNII yields, we use \citet{Nomoto-06}, parameterised as a
function of metallicity; we interpolate the yields to the metallicity
of the given gas element being enriched.  We are using essentially
the same yield tables used in \citet{Oppenheimer-08}, which in
\citet[and other studies]{Dave-11b} was found to result in metallicities
that are too high by a factor of $\sim 2$ compared to the observed
mass-metallicity relation.  In previous
studies~\citep[e.g.][]{Dave-13,Liang-16} we have simply rescaled
metallicities in post-processing, arguing that such changes do not
greatly impact the dynamics, and that the predicted yields are
uncertain at this level.  However, given that cooling and $H_2$-based
star formation rates depend on metallicity, it is more accurate to
reduce the yields during the simulation run.  For this reason, we
multiply all Type~II SN yields by a factor of 0.5.  We add all
SNII energy to the star-forming gas element itself, in the
instantaneous self-enrichment approximation.  On cosmological
timescales, and with sufficient star-forming gas within a given
galaxy, this yields a well-converged metallicity distribution.

For SNIa yields, we employ \citet{Iwamoto-99}, assuming each SNIa
yields $1.4 M_\odot$ of metals.  The prompt SNIa component, like
TypeII enrichment, is done in the instantaneous self-enrichment
approximation, while the delayed feedback components are added to
the 16 nearest gas element surrounding the given star in a
kernel-weighted manner.

For AGB stars, we follow \citet{Oppenheimer-08} and employ a lookup
table as a function of age and metallicity in order to obtain the
yields; these yields come from several sources, as described in
\citet{Oppenheimer-08}.  We additionally assume 36\% helium fraction
and a nitrogen yield of 0.00118.  The enrichment, like the energy,
is added from stars to the nearest 16 gas elements, kernel-weighted,
following the mass loss rate as computed assuming a \citet{Chabrier-03}
IMF.

For the KMT model, we need to provide a total metallicity in solar
units, for which we assume a solar abundance taken from \citet{Asplund-09}.
We do not include metal diffusion or mixing in these models to avoid
an extra free parameters, since it is unclear how much metal mixing
should be occuring~\citep{Schaye-07}.

%

\begin{table*}
\caption{$2\times 512^3$-particle \mufasa\ runs, with reduced $\chi_\nu^2$ values relative to observed GSMFs at $z=0-6$.}
\begin{tabular}{lccccccccccc}
\hline
Name &
$L$ &
$\epsilon$ &
$m_{\rm gas}$ &
$m_{\rm dark}$ &
$m_{\rm gal}$ &
$\chi^2_{z=0}$ &
$\chi^2_{z=1}$ &
$\chi^2_{z=2}$ &
$\chi^2_{z=3}$ &
$\chi^2_{z=4}$ &
$\chi^2_{z=6}$\\
\hline
\multicolumn {4}{c}{} \\
m50n512   & 50 & 0.5 & $1.82\times 10^7$ & $9.6\times 10^7$ & $5.8\times 10^8$      & 0.20 & 0.98 & 1.11 & 0.23 & 1.32 & 1.97 \\
m25n512  & 25 & 0.25 & $2.28\times 10^6$ & $1.2\times 10^7$ & $7.3\times 10^7$       & --- & ---- & 1.04 & 0.08 & 1.24 & 1.00 \\
m12.5n512 & 12.5 & 0.125 & $2.85\times 10^5$ & $1.5\times 10^6$ & $9.1\times 10^{6}$ & --- & ---- & 0.19 & 0.02 & 0.38 & 0.26   \\
\hline
\end{tabular}
\\
$L$ is the box size in comoving $\hmpc$.\\
$\epsilon$ is the minimum gravitational softening length in comoving $\hkpc$ (equivalent Plummer).\\
$m_{\rm gal}$ is the galaxy stellar mass resolution limit in $M_\odot$ (32 star particle masses).\\
\label{tab:sims}
\end{table*}

\subsection{Quenching Feedback}

With only stellar feedback as described above, massive galaxies
still form stars too rapidly and the stellar mass function is not
truncated at the high mass end~\citep[e.g.][]{Gabor-10}.  Hence an
additional ``maintenance mode" source of energetic feedback is
required to fully quench massive
galaxies~\citep[e.g.][]{Croton-06,Somerville-08}.  Here we follow
the model in \citet{Gabor-12,Gabor-15}, who argued that keeping gas
hot in massive halos is sufficient to yield quenched massive galaxies
in agreement with mass function and red sequence observations.

We quench galaxies by heating all gas in massive halos {\it except} gas that
is self-shielded.  We compute the self-shielded cold gas fraction for
each gas element using the prescription described in \citet{Rahmati-13},
which parameterises the effective strength of the photo-ionising background
impinging on gas as a function of density and redshift.  This
ionising background strength is then used to compute the cold gas
fraction (which includes both neutral and molecular gas) using a
rate balance equation as described in \citet{Popping-09}.  Gas is
considered self-shielded if its cold gas fraction after applying
self-shielding is above 10\%.  By only heating non-self-shielded
gas, most of the neutral and molecular hydrogen within hot halos
is unaffected, ensuring that these cold gaseous components are only
directly affected by physical processes such as stripping rather
than our quenching prescription.

We apply this quenching heating in massive halos with a mass that
exceeds the quenching mass given by
\begin{equation}
M_q = (0.96 + 0.48z)\times 10^{12}M_\odot.
\end{equation}
This formula is obtained from the analytic equilibrium model of
\citet{Mitra-15} as the best-fit parameterization of $M_q$ required
to match various observational constraints from $z=0-2$.

Gas in halos above $M_q$ that is not self-shielded and is below the
virial temperature $T_{\rm vir}$ is heated to 20\% above virial
temperature $T_{\rm vir}$ at each timestep.  We calculate $T_{\rm
vir} = 9.52\times 10^7 M_h^{1/3}$K~\citep{Voit-05}, where we obtain
the halo mass $M_h$ using a friends-of-friends algorithm with a
linking length of 0.16 times the mean interparticle spacing.  This
FOF algorithm is different than the one used to compute $M_*$ and
$v_c$, and is only run every 0.5\% of a Hubble time only on dark
matter particles since it has substantial computational cost; gas
and star particles are linked to halos via their nearest dark matter
particle.

We stress that this is a phenomenological model intended to capture
the most important effects of whatever microphysics actually drive
quenching.  Specifically, we are {\em not} saying that halos simply
``self-quench'' in the simulations (owing to e.g.\ virial shocks
or simple cooling physics). Quite the opposite: if we only include
hydrodynamics, cooling, star formation, and stellar feedback physics,
we see far too much star formation at high masses, as noted above.
This model directly requires some additional feedback mechanism to
be the source of (considerable) additional energy to maintain the
virial temperatures of halo gas. The physical mechanisms of this
feedback remain uncertain, however -- hence the empirical model
adopted here. The most likely candidate is feedback from a super-massive
black hole, which many groups have shown can plausibly inject
sufficient energy far out in the halo and resolve the cooling flow
problem~\citep{Croton-06,Somerville-08,Schaye-15}. In future work,
we will explore what the empirical model here requires in terms of
energetic input, and how this compares to e.g. the kinetic luminosity
function of AGN jets. In these simulations (with $\sim$kpc resolution),
however, any model for black hole formation/seeding, dynamical
evolution, merging, fueling/growth, and feedback would be necessarily
sub-resolution. As such, we would in any case simply adjust the
sub-resolution parameters until they produced the same effects as
our quenching model here. Our method here thus allows us to more directly
isolate what needs to occur on {\em resolved} scales in the simulation,
in order to match massive galaxy observations.

\subsection{Runs}\label{sec:runs}

All simulations assume a cosmology consistent with \citet{Planck-15}
``full likelihood" constraints:  $\Omega_m=0.3$, $\Omega_\Lambda=0.7$,
$\Omega_b=0.048$, $H_0=68\kmsmpc$, $\sigma_8=0.82$, and $n_s=0.97$.
We generate initial conditions with these parameters using {\sc
Music}~\citep{Hahn-11}.  Each runs begins at $z=249$ in the linear
regime.

Our fiducial simulation consists of a $50\hmpc$ randomly-selected
cubical cosmological volume, with additional simulations used to
examine early galaxy evolution having $25\hmpc$ and $12.5\hmpc$ box
sizes (these are separate runs, not sub-volumes extracted from the
larger run).  Each run evolves $512^3$ gas fluid elements (i.e.
mass-conserving cells) and $512^3$ dark matter particles.  With
these choices, the $12.5\hmpc$ volume's halo mass resolution (i.e.
64 dark matter plus 64 gas elements) is $\approx 10^8 M_\odot$,
which means that this run resolves essentially all hydrogen cooling
halos back to the earliest epochs.  We will also show results from
$25\hmpc$, $2\times 256^3$ element runs to examine variations in
hydrodynamics methodology.

Table~1 lists our main \mufasa\ run parameters.  The 50$\hmpc$ boxes
is evolved to a final redshift of $z_{\rm end}=0$, while the 25$\hmpc$
and $12.5\hmpc$ volumes down to $z_{\rm end}=2$ owing to both
computation expense and the fact that these volumes are too small
to be representative at later epochs.  We also list the galaxy
stellar mass resolution limit $M_{\rm *,min}$ in each run, taken
to be 32 star particle (or equivalently, gas element) masses.  We
have found that stellar mass functions are generally well-converged
to this limit, but other quantities such as star formation histories
can require a higher mass threshold for
convergence~\citep[e.g.][]{Dave-11a}.

We output 135 snapshots down to $z=0$ (85 to $z=2$) for each run.
We analyse the snapshots using SPHGR\footnote{\tt
http://sphgr.readthedocs.org/en/latest/} \citep{Thompson-15}, which
identifies galaxies using SKID and halos using {\sc
RockStar}~\citep{Behroozi-13}, and then links them via their
positions.  SPHGR goes on to calculate many basic properties of the
galaxies and halos such as $M_*$ and SFR, and finally outputs a
Python ``pickle" file for each snapshot that contains all this
information, including member particle lists.  All plots in this
paper are produced directly from these SPHGR pickle files.

\section{Stellar Mass Functions}\label{sec:gsmf}

\begin{figure*}
  \centering
  \subfloat{\includegraphics[width=0.5\textwidth]{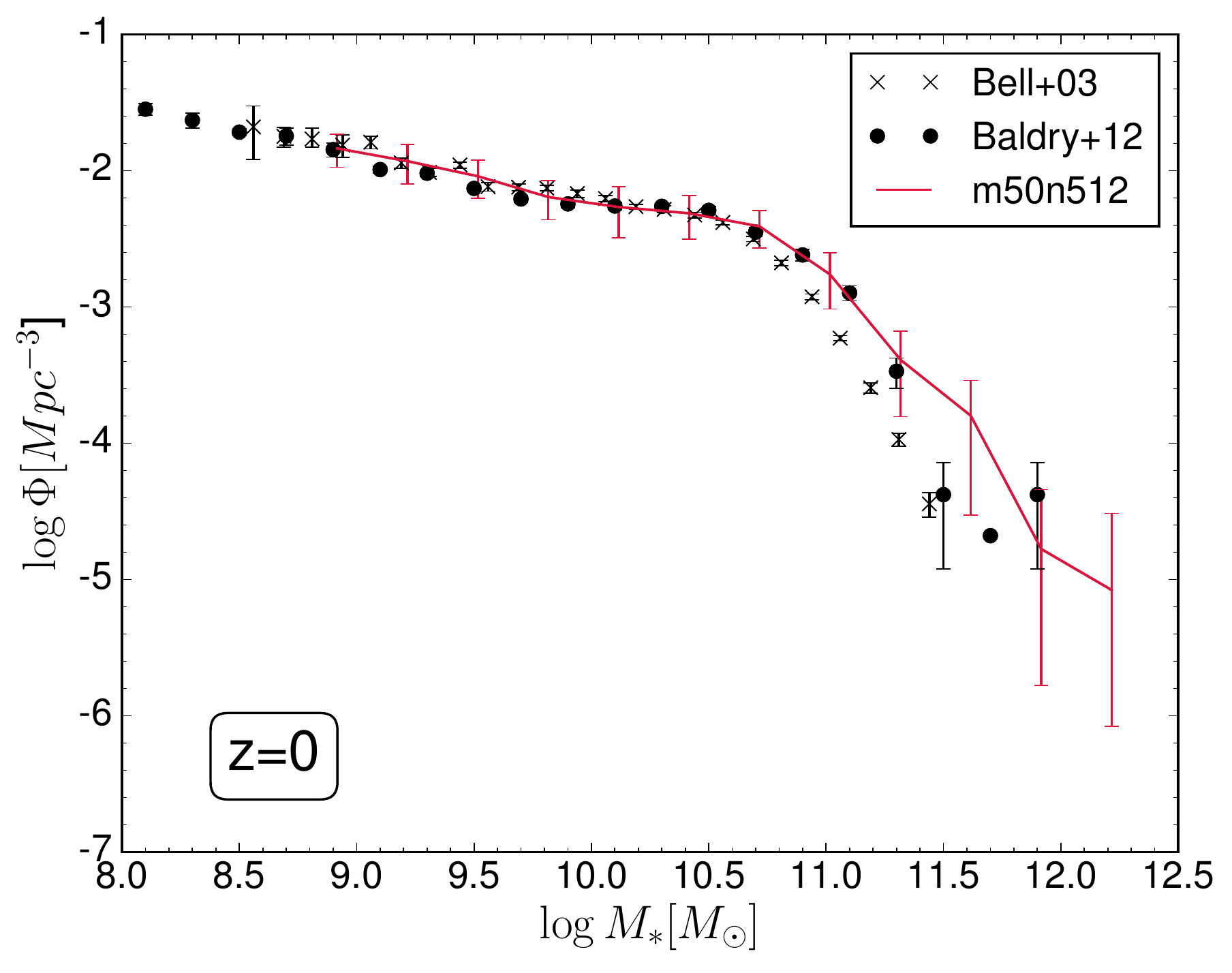}}
  \subfloat{\includegraphics[width=0.5\textwidth]{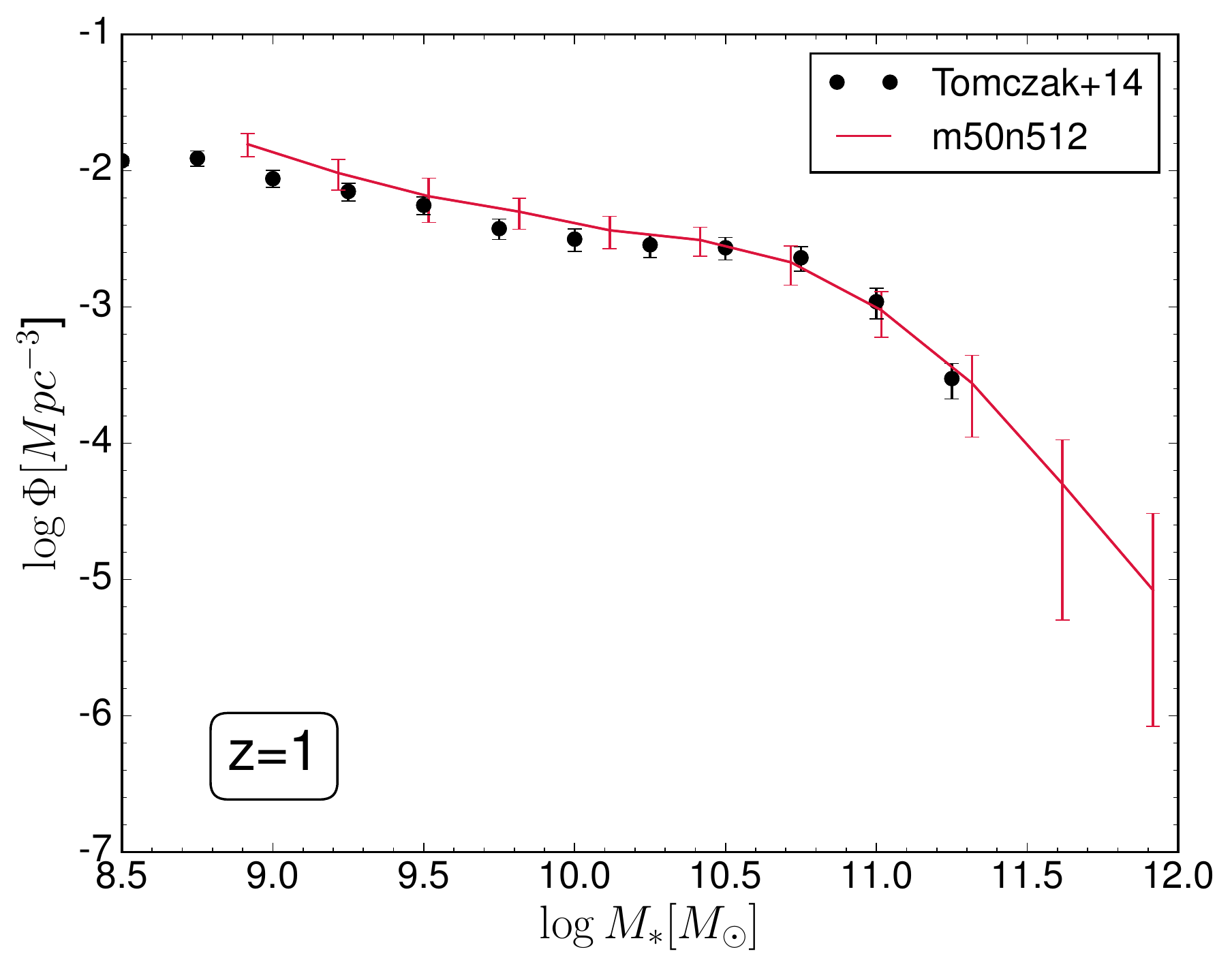}}
  \hfill
  \vskip-0.1in
  \subfloat{\includegraphics[width=0.5\textwidth]{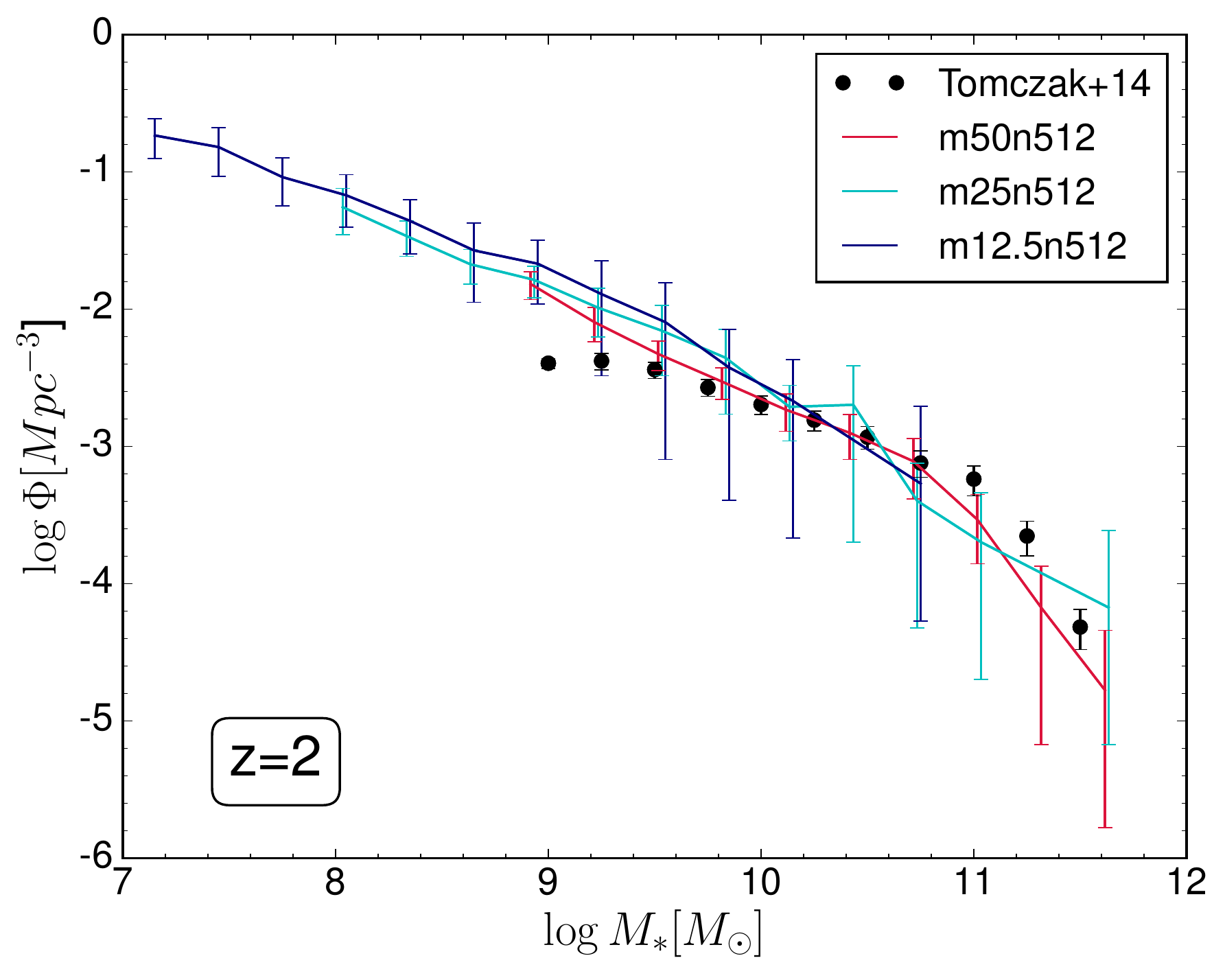}}
  \subfloat{\includegraphics[width=0.5\textwidth]{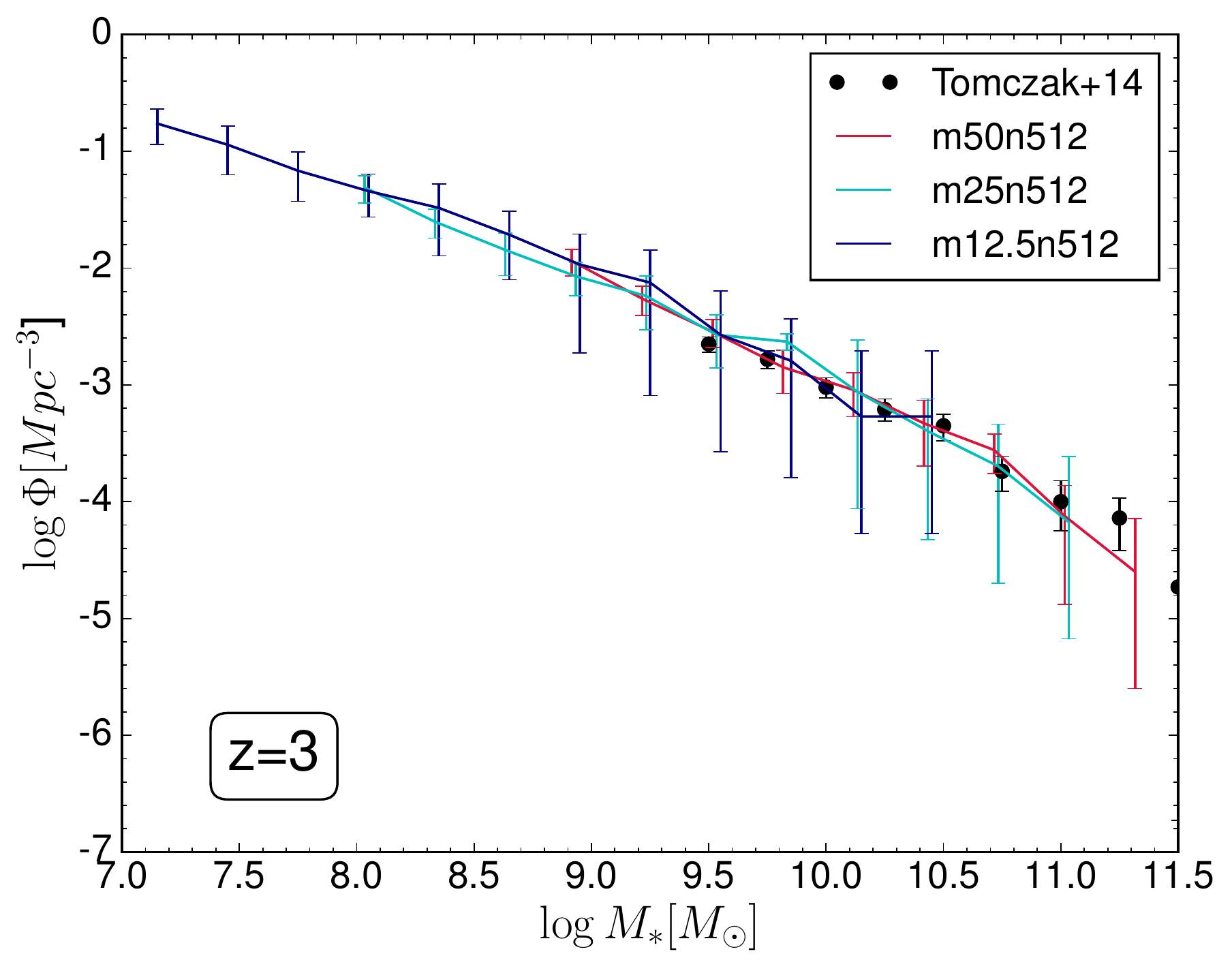}}
  \hfill
  \vskip-0.1in
  \subfloat{\includegraphics[width=0.5\textwidth]{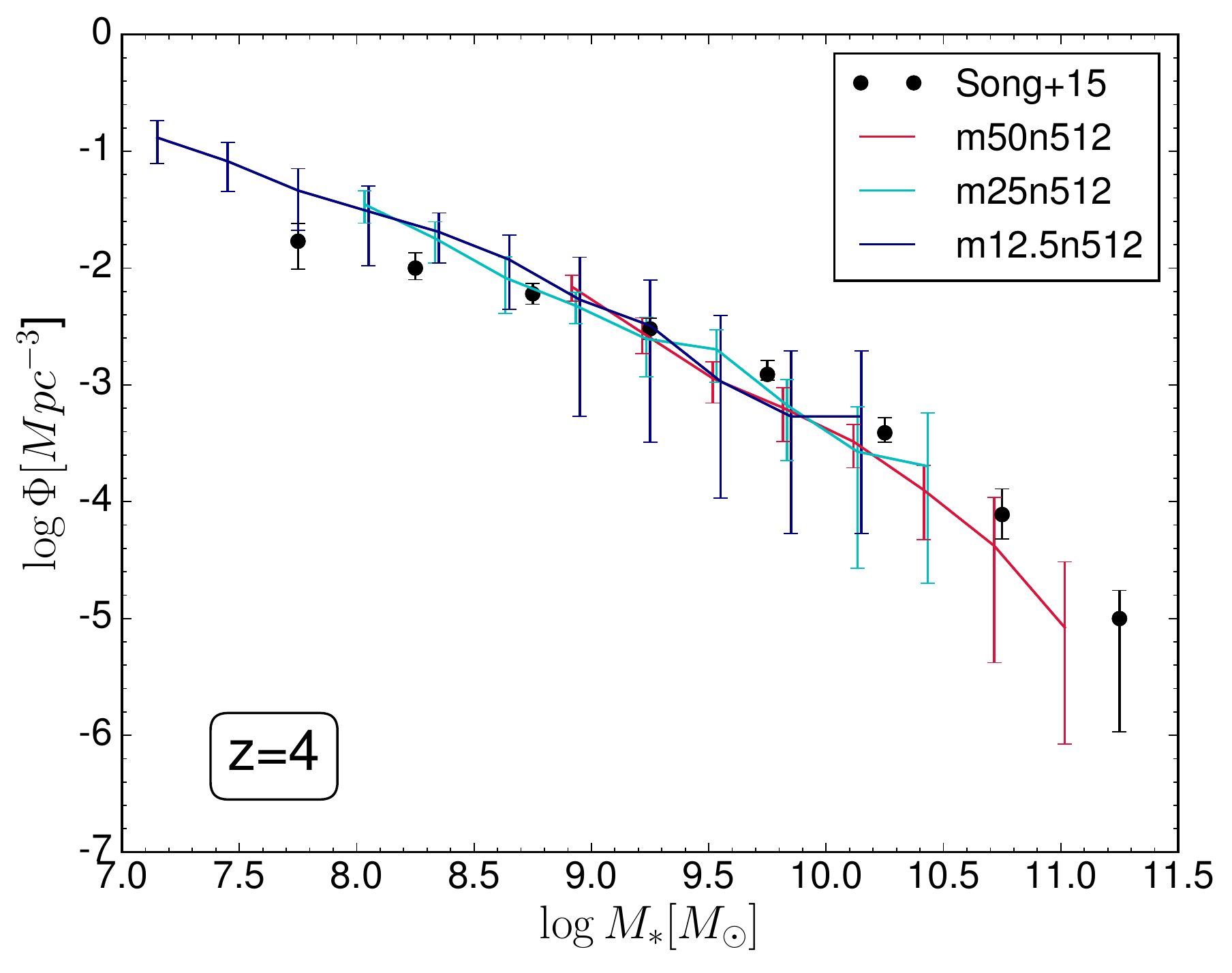}}
  \subfloat{\includegraphics[width=0.5\textwidth]{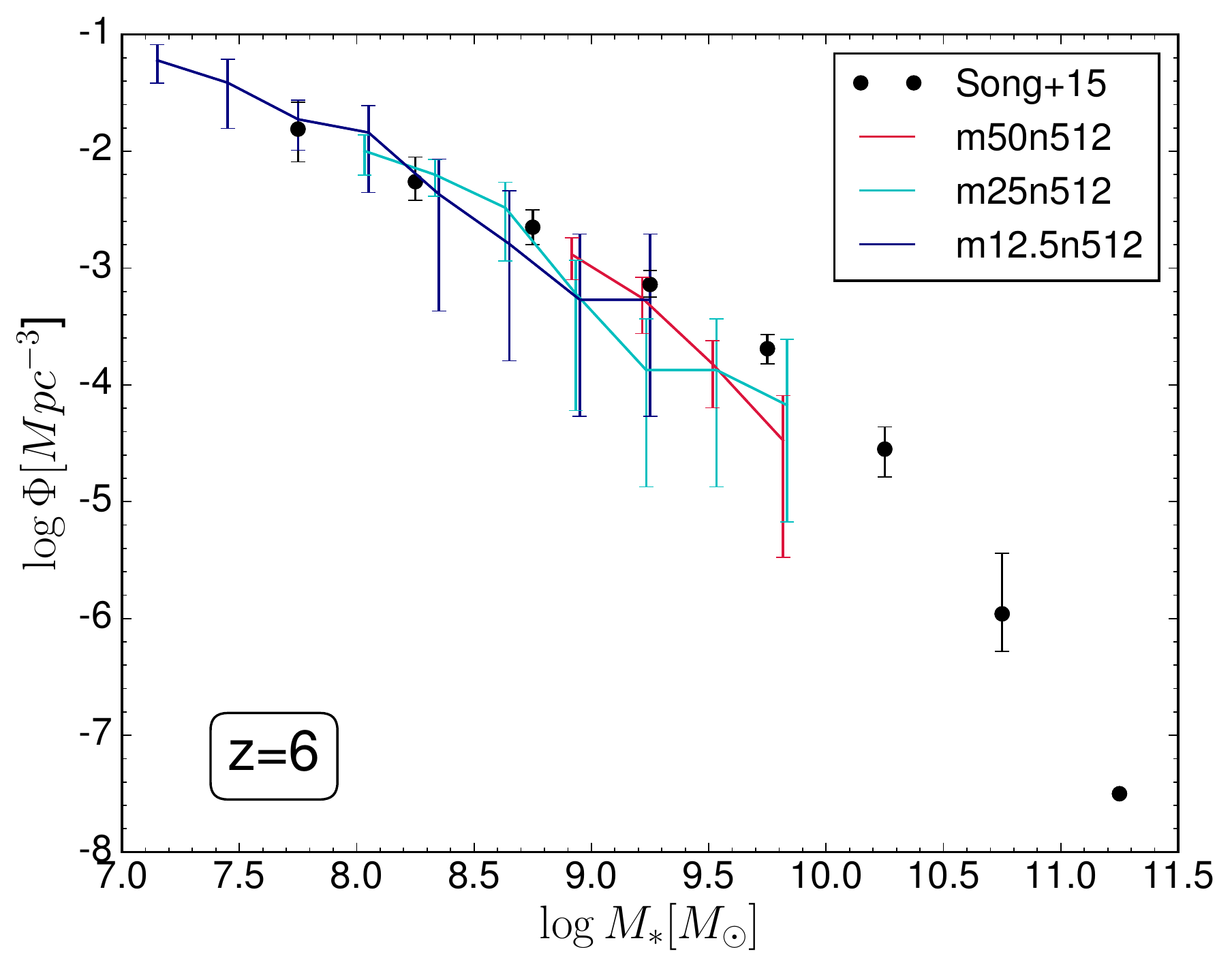}}
  \vskip-0.1in
  \caption{ Galaxy stellar mass functions at $z=0,1,2,3,4,6$.  Dark red, cyan, and dark blue 
curves show the results from our $50, 25,$ and $12.5\hmpc$ volumes, respectively.
Simulated galaxies are included down to the stellar mass resolution limits listed in 
Table~\ref{tab:sims}.  Uncertainties are computed from cosmic variance over 8
sub-octants of each simulation volume.  Observations are shown as solid points from 
\citet{Baldry-12} at $z\sim 0$, \citet{Tomczak-14} at $z=1-3$, and \citet{Song-15} at $z=4,6$.
}
\label{fig:mfres}
\end{figure*}

The primary observational benchmark for testing galaxy formation
models is the galaxy stellar mass function (GSMF).  The GSMF has
now been measured robustly out to $z\sim 3$, covering about 90\%
of cosmic stellar mass buildup, thanks in particular to improving
deep near-infrared surveys such as the Cosmic Assembly Near-IR Deep
Extragalactic Legacy Survey~\citep[CANDELS;][]{Grogin-11,Koekemoer-11}
and the FourStar Galaxy Evolution Survey~\citep[ZFOURGE;][]{Tomczak-14}.
Measurements at higher redshifts are also improving
rapidly~\citep[e.g.][]{Duncan-14,Song-15}, though they typically
rely on less certain conversions between ultraviolet luminosity and
stellar mass.  Matching the GSMF and its evolution is thus a key
test of whether a model reproduces the bulk of galaxy growth over
cosmic time.

It remains a significant challenge for galaxy formation models, be
they semi-analytic (SAMs) or hydrodynamic, to match the GSMF to
within uncertainties at all redshifts~\citep{Somerville-15}.  SAMs
can reproduce the $z=0$ GSMF quite well, typically because they are
tuned to do so, while hydrodynamic simulations have recently achieved
improved levels of success with $\la\times 2$ discrepancies at
$z=0$.  Moving to higher redshifts, most models overproduce the
GSMF below the knee at $z\sim 1-2$, and some underproduce the GSMF
at $z\ga 4$.  These differences show up prominently when considering
the star formation histories of galaxies, where models characteristically
deviate from that inferred for real galaxies particularly at lower
masses in the sense of having overly early stellar growth in low-mass
galaxies~\citep{White-15}.  In this section, we will test the
\mufasa\ simulations against observations using the evolution of
the GSMF from $z=6\rightarrow 0$.

\subsection{Comparisons to observations}

Figure~\ref{fig:mfres} shows the GSMF from our \mufasa\ simulations
at $z=0,1,2,3,4,6$.  The three box sizes are colour-coded as navy
blue ($12.5\hmpc$), cyan ($25\hmpc$), and crimson red ($50\hmpc$).
Only galaxies with stellar masses above the galaxy mass resolution
limit listed in Table~\ref{tab:sims} are included in each case.  We
compare to observations (black circles, with Poisson errors) at
$z=0$ from the Galaxy and Mass Assembly (GAMA) Survey by
\citet{Baldry-12} and from \citet{Bell-03}, at $z=1,2,3$ to combined CANDELS and ZFOURGE
data from \citet{Tomczak-14}, and at $z=4,6$ to CANDELS data from
\citet{Song-15}.  For \citet{Tomczak-14}, to obtain the GSMF at
$z=1$ we average the quoted values at $z=0.75-1$ and $z=1-1.25$
weighted by the number of galaxies in each bin, and analogously we
compute a weighted average of $z=1.5-2$ and $z=2-2.5$ to obtain the
$z=2$ points shown; for $z=3$, we quote the $z=2.5-3$ results, which
is likely a slight overestimate.  While we only choose to compare
to one data set so we can more easily engage in quantitative
statistical comparisons, various current GSMF observations do not
differ significantly at $z\la 2$ \citep[e.g.][]{Somerville-15}.

In Table~1, we quote a reduced $\chi^2$ value comparing the simulation
data to observational data at each redshift, for each run.  To
obtain this, we compute error bars on the simulation values as the
variance in the GSMF over eight sub-octants of each simulation
volume.  This approximately accounts for both cosmic variance and
Poisson errors, though generally the Poisson errors are sub-dominant.
This then produces error bars (asymmetric in log space), which are
displayed in Figure~\ref{fig:mfres}.  We then compute $\chi^2$ from
these error bars in log (not linear) space, so that a GSMF that is
e.g.  low by a factor of 2 will have the same $\chi^2$ as one that
is high by a factor of 2.  We sum $\chi^2$ over the range of
overlapping masses between the simulation and the data, using the
upper error bar if the data is higher than the simulation, otherwise
the lower error bar.  To get a reduced $\chi^2_\nu$ we divide by
the number of observed data points.  These $\chi_\nu^2$ values
provide a quantitative assessment of the level of agreement between
\mufasa\ runs and the observed GSMF.

We note that this value of $\chi^2$ does not take into account
uncertainties in the observations themselves, but as can be seen
in Figure~\ref{fig:mfres} from the statistical error bars shown on
the observed points, these uncertainties are generally sub-dominant
to cosmic variance uncertainties.  There are additional systematic
errors in observed GSMF determinations which can be quite significant
particularly at high redshifts, but we also ignore these since these
are difficult to quantify accurately~\citep[see e.g.][for more
discussion]{Mobasher-15}.  We also note that we are comparing
SKID-derived masses in simulations to (typically) aperture-based
masses from various observational samples, which may introduce a
significant bias particularly at high masses ($M_*\ga 10^{11}M_\odot$)
where high-Sercic profiles create extended wings in the surface
brightness distribution~\citep{Genel-14,Kravtsov-14}.  Hence the
comparisons at the highest masses should be regarded with some
additional observational uncertainty, but our values for $\chi_\nu^2$
are generally driven by low-$M_*$ where our cosmic variance errors
are smaller.  Finally, we note that our cosmic variance errors may
still be underestimated owing to our smaller volume, as \citet{Genel-14}
found somewhat larger uncertainties from 25$\hmpc$ sub-volumes in
Illustris.  All of these additional aspects would go towards lowering
our quoted $\chi^2_\nu$ values.

We now examine how well the mass functions agree with observations
at various redshifts.  At $z=0$, our $50\hmpc$ fiducial \mufasa\
run is in remarkably good agreement with the \citet{Baldry-12} GSMF.
It faithfully reproduces the observed low-mass slope down to our
galaxy mass resolution limit, showing that our star formation
feedback algorithm does very well at properly suppressing the
low-mass end.  At high masses, there is a sharp turnover towards
an exponential that has been difficult to achieve in previous models,
and highlights the success of translating the evolving quenching
mass determined from the equilibium model of \citet{Mitra-15} into
full hydrodynamic simulations.  There may be a very slight
overproduction of the highest mass galaxies, albeit well within
cosmic variance uncertainties, but we note that there is some
uncertainty in the masses of massive ellipticals owing to how the
galaxy is separated from intracluster light~\citep{Kravtsov-14}.
As listed in Table~\ref{tab:sims}, the reduced $\chi^2_\nu=0.20$
at this redshift relative to \citet{Baldry-12}, which quantifies
the excellent agreement with observations.  The agreement is slightly
less good veresus \citet{Bell-03}, with $\chi^2_\nu=0.58$, driven mostly
by the reduced high-mass GSMF in this determination.

At $z=1$, the \mufasa\ mass function is in good agreement with the
\citet{Tomczak-14} observations, with $\chi_\nu^2\approx 1$.  There
is a slight excess in the sub-$M^\star$ range; this becomes more
prominent at $z=2$, which has a comparable $\chi_\nu^2\approx 1.1$.
The disagreement is driven most strongly by the lowest observed
point, which may also be the most uncertain owing to incompleteness
in the observations.  In contrast, \citet{Reddy-09} determined a
steeper faint-end slope of $-1.7$ from rest-UV observations, in
much better agreement with \mufasa.  Upcoming observations with the
{\it James Webb Space Telescope} should definitively settle this
issue.

At $z=2$ we see, in both the observations and the $50\hmpc$ simulation,
a turn-down at the highest masses where our quenching model starts
to suppress massive galaxy growth.  At this redshift, $M_q\approx
10^{13}M_\odot$, which means that galaxies with $M_*\ga 10^{11}M_\odot$
can begin to be affected.  The onset of quenched massive galaxies
between $z\sim 2-3$ is in general agreement with
observations~\citep[e.g.][]{Kriek-09}.  Interestingly, the most
actively star-forming galaxies in the universe at $z\sim 2.5$ have
halo masses of $\sim 10^{13-13.5}M_\odot$~\citep{Wilkinson-16},
which corroborates our value of $M_q$ at these epochs.

Since our smaller volumes are evolved to $z=2$, we can here examine
the resolution convergence among all three volumes.  Broadly, the
three runs lie within cosmic variance uncertainties of each other.
However, in detail we see that there is a systematic trend that
higher resolution simulations produce slightly higher mass functions
in the overlapping mass range.  Over a factor of 64 in mass resolution
between the $50\hmpc$ and $12.5\hmpc$ volumes, the mass function
at $M_*\approx 10^{9-9.5}$ is increased by about a factor of two.
At higher masses, the GSMF in the smallest volume start to drop low
because it cannot form enough massive galaxies.  The $25\hmpc$
interpolates this trend, being slightly higher at low-$M_*$ compared
to the fiducial volume, and slightly lower at high-$M_*$.  We note
that the $\chi^2_\nu$ values for the smaller volumes are still close
to unity, so the agreement with data is still satisfactory, although
in part this owes to our smaller volumes having larger cosmic
variance uncertainties.

At $z=3$, all the volumes' GSMFs are essentially power laws over
the modeled mass range.  The agreement is excellent with all three
simulations, with $\chi^2_\nu\la 0.3$.  However, we note that the
data used is actually from galaxies at $z=2.5-3$, so at $z=3$ it
is possible that the data should be slightly lower than indicated.
Hence as a check we analyzed a $z=2.7$ simulation output and compared
with this data, and still found very good agreement with $\chi^2_\nu\la
0.5$.  The resolution convergence among the three volumes here, as well
as at all higher redshifts, is excellent.

At higher redshifts ($z=4,6$), we begin to see slight discrepancies
at the high-mass end, in the sense that the \mufasa\ runs do not
produce enough high-mass galaxies.  It is already slightly evident
at $z=4$ -- while the smallest volume has $\chi_\nu^2<1$, the larger
volumes show $\chi_\nu^2=1.3$.  The discrepancy becomes somewhat
more significant at $z=6$, particularly in the $50\hmpc$ volume,
where $\chi_\nu^2\approx 2$.  The other volumes, probing only lower
masses, are not as strongly discrepant, having $\chi^2_\nu\leq 1$
at $z=6$.  This discrepancy could be the result of the very high
bias of early massive galaxies, which reside in large halos that
are not properly represented even in the $50\hmpc$ volume.
Alternatively, it could be that the limited resolution does not
capture very early star formation in the largest density perturbations.
Hence both resolution and volume effects go towards underproducing
early stellar mass.  We note that another high-$z$ GSMF determination
from CANDELS by \citet{Duncan-14} yields generally higher values,
which would be further discrepant; the determination from
\citet{Grazian-15}, on the other hand, are comparable to that of
\citet{Song-15}.  Nonetheless, it is worth noting that are at present
substantial systematic uncertainties in deriving stellar masses at
these redshifts.

We note that at $z\ga 2$, and particularly at $z\ga 4$, the
observational GSMF determinations are subject to significant
systematic uncertainties, and various determinations often lie
significantly outside each others' formal statistical
uncertainties~\citep{Ilbert-13,Muzzin-13,Tomczak-14,Mobasher-15}.
For this reason, one should use caution in interpreting the agreement
or lack therefore with models at these redshifts, as improving
observations may shift these values non-trivially.  Nonetheless,
much of the disagreement between data sets occurs at high-$M_*$,
whereas our $\chi^2_\nu$ is driven by our agreement at low-$M_*$
where our cosmic variance errors are small.  For now, we claim that
our agreement is comparably good against most recent data sets at
these redshifts, but that future observational determinations may
yet alter this.

In summary, \mufasa\ does an excellent job of matching the observed
GSMF from $z=4\rightarrow 0$, in all cases to within $\la 1.2\sigma$
in simulation cosmic variance.  This agreement is unprecedented,
and suggest that \mufasa\ achieves an impressive level of realism
in modeling galaxy stellar growth across cosmic time.  The agreement
at the massive end owes to our evolving halo quenching mass, and
suggests that this simple description provides a good match to the
truncation of star formation across cosmic time~\citep[as argued
e.g. in][]{Gabor-12,Gabor-15}.  It also illustrates the synergy
between constraints derived from the analytic equilibrium model and
detailed hydrodynamic simulations.  At high redshifts, the data are
still subject to significant systematic uncertainties, hence we
await upcoming deeper observations particularly using the {\it James
Webb Space Telescope} to more robustly constrain these models.
Nonetheless, our high-resolution simulations agree quite well with
currently available reionisation-epoch GSMF observations, suggesting
that simulations that resolve all H-cooling halos are robust for
such studies.  Our runs also show reasonable resolution convergence,
albeit with a noticeable trend for higher-resolution simulations
to produce slightly higher GSMFs at the low-mass end.

\subsection{Comparison to other simulations}

\begin{figure*}
  \centering
  \subfloat{\includegraphics[width=0.5\textwidth]{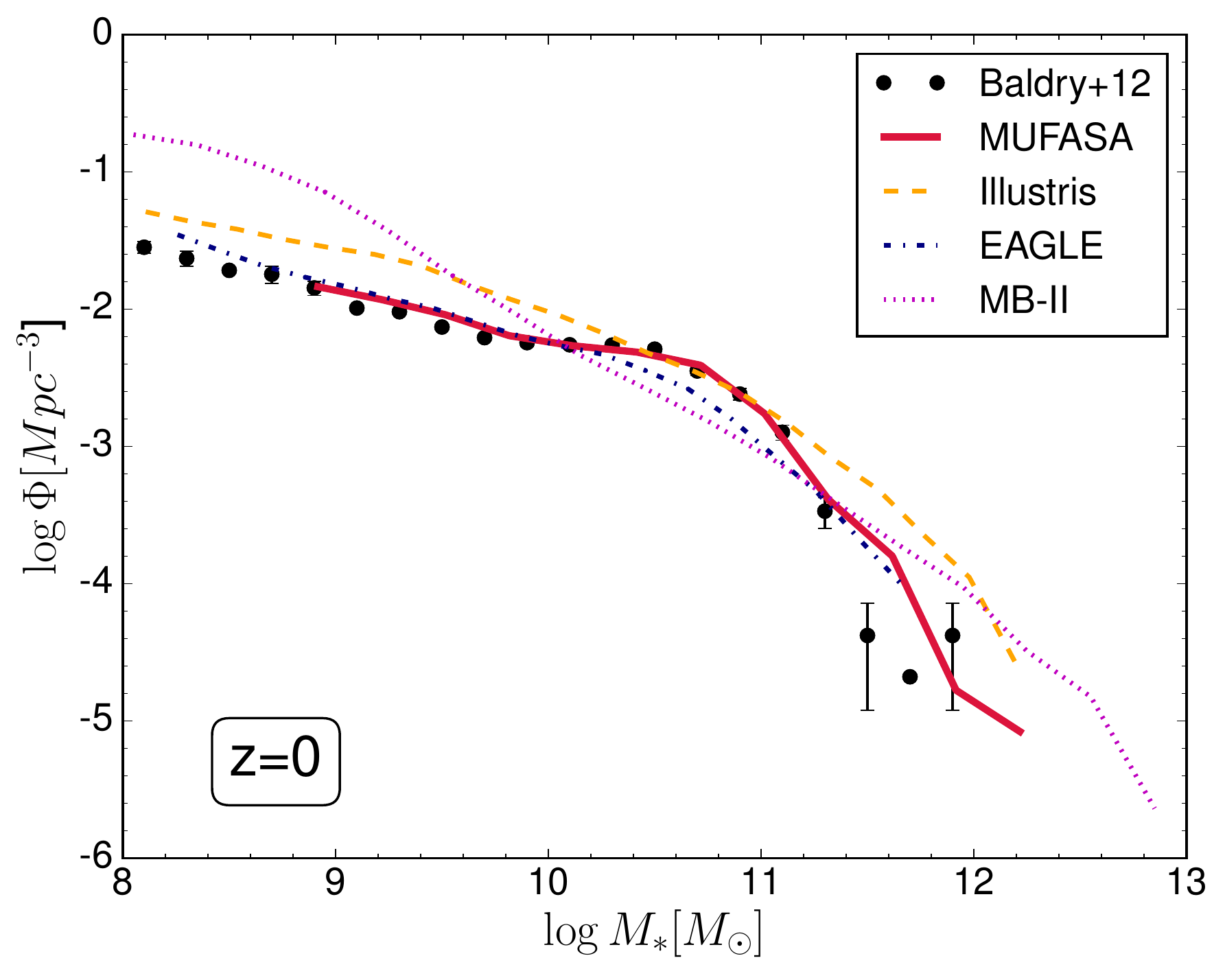}}
  \subfloat{\includegraphics[width=0.5\textwidth]{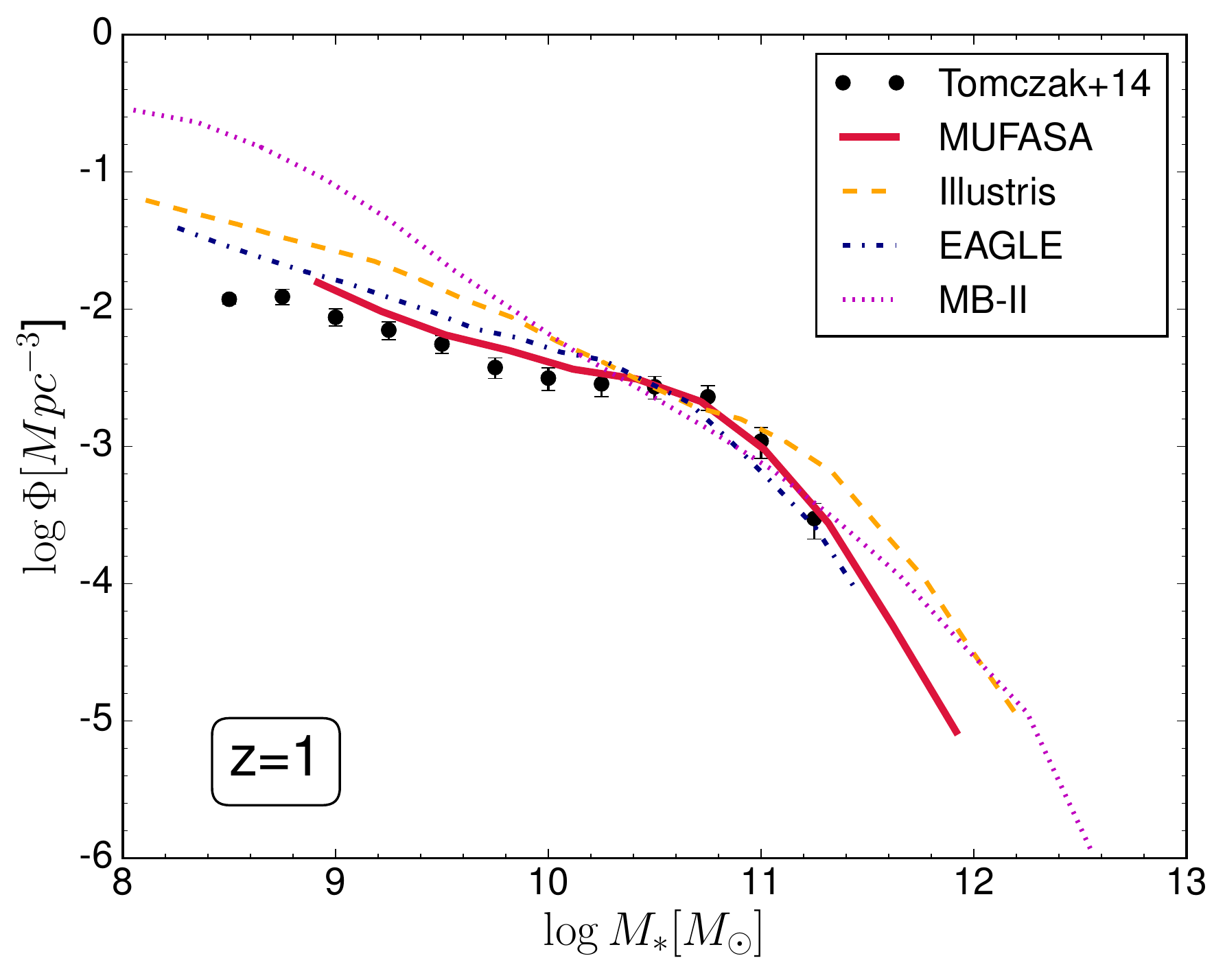}}
  \hfill
  \vskip-0.2in
  \subfloat{\includegraphics[width=0.5\textwidth]{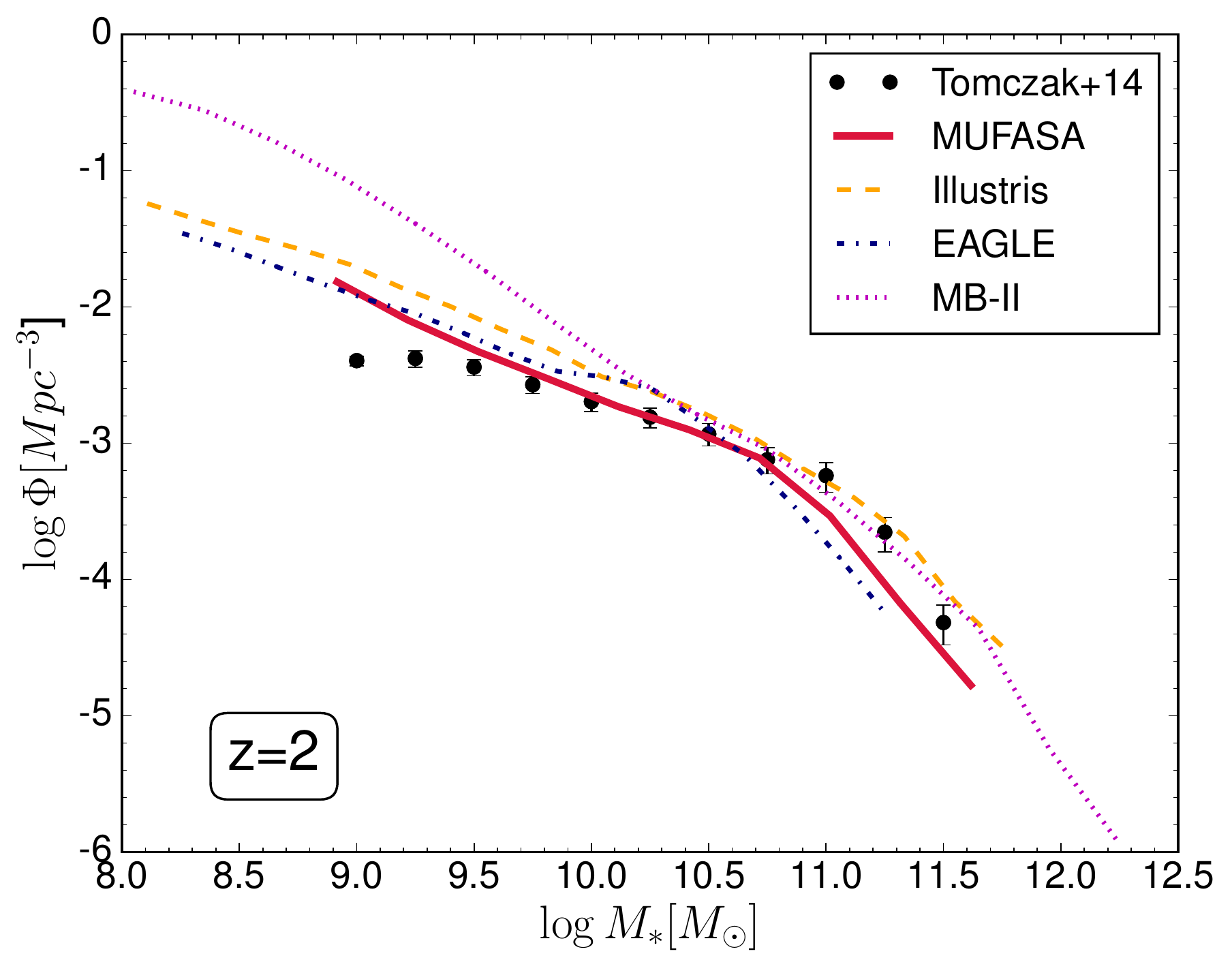}}
  \subfloat{\includegraphics[width=0.5\textwidth]{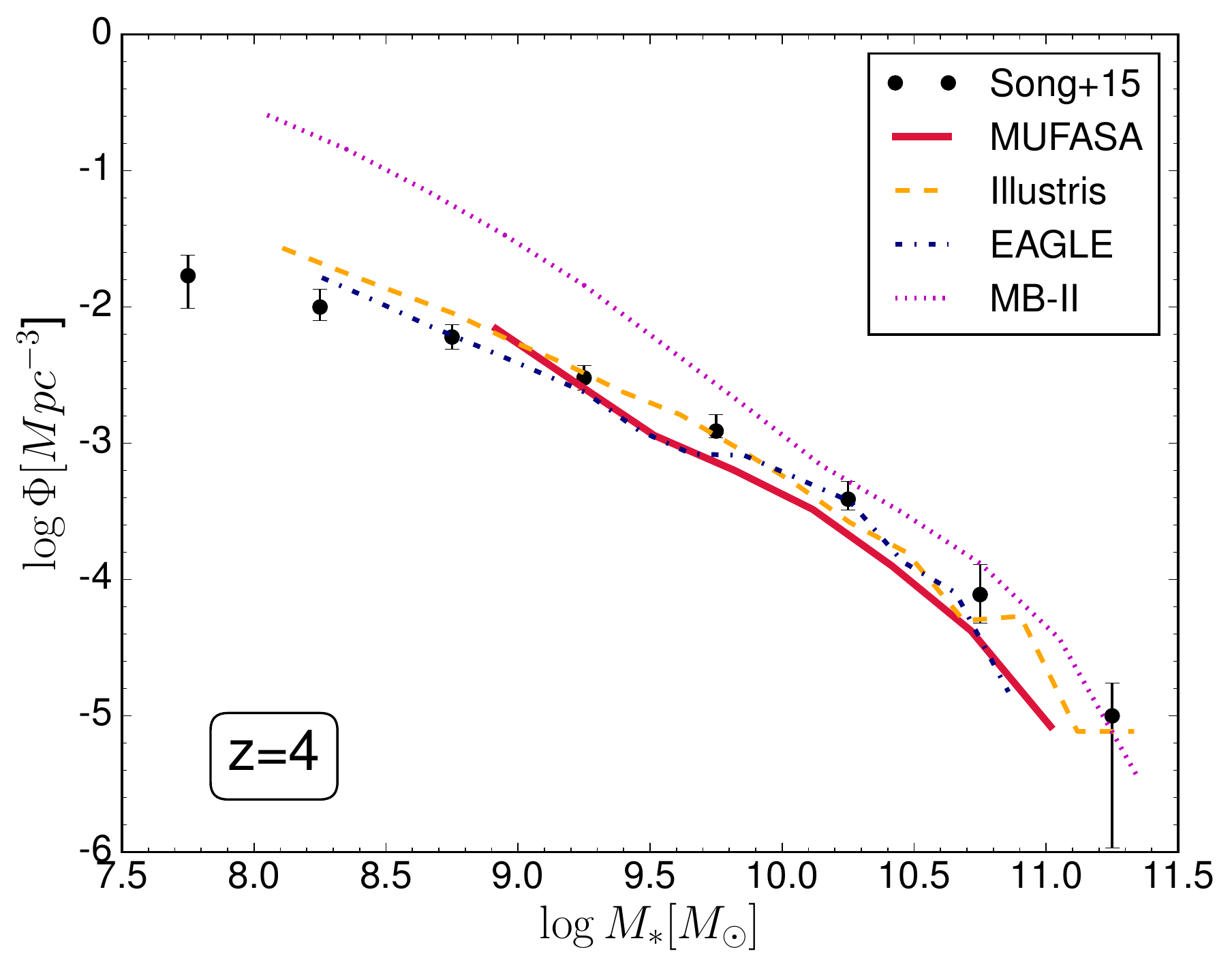}}
  \hfill
  \vskip-0.1in
  \caption{ Galaxy stellar mass functions at $z=0,1,2,4$ as in Figure~\ref{fig:mfres},
with uncertainties omitted for clarity and the same colour scheme as before,
now compared to results from the Illustris~\citep[orange dashed;][]{Vogelsberger-14}, 
EAGLE~\citep[dark blue dashed;][]{Schaye-15}, and MassiveBlack-II~\citep[magenta dashed;][]{Khandai-15} simulations.  Observations are shown as described in Figure~\ref{fig:mfres}.  
\mufasa\ produces better agreement with observations than other current models from $z\sim 0-2$,
and slightly worse at $z\sim 4$.}
\label{fig:mfsim}
\end{figure*}

Recently, major cosmological galaxy formation simulation projects
have set new standards for the number of fluid elements and input
physics.  Illustris~\citep{Vogelsberger-14,Genel-14},
EAGLE~\citep{Schaye-15,Crain-15}, and
MassiveBlack-II~\citep[MB-II;][]{Khandai-15} each model around
$1800^3$ fluid elements, making them the (computationally) largest
cosmological hydrodynamic simulations of galaxy formation evolved
to $z=0$ to date.  Illustris employs {\sc Arepo}, EAGLE uses a
modern SPH variant, and MB-II uses \gad\ with entropy-conserving
SPH~\citep{Springel-05}.  While the dynamic range of these simulations
is a remarkable computational achievement, it is still insufficient
to directly model the feedback processes that regulate galaxy
evolution.  Hence like \mufasa, they utilize subgrid representations
for star formation and black hole feedback in order to suppress
galaxy growth.  The differences in their GSMF predictions are likely
driven primarily by variations in feedback recipes, as opposed to
variations in hydrodynamic methodology~\citep[see e.g.] [and
\S\ref{sec:hydro}]{Hopkins-13}.

Illustris and MassiveBlack-II employ decoupled winds, phenomenologically
similar to \mufasa.  Illustris uses scalings for energy-driven
winds, namely $v_w\propto v_c^{-2}$; this is similar to that predicted
from high-resolution zooms by \citet{Christensen-16}, although the
amplitude assumed in Illustris is many times higher.  MB-II
uses a constant mass loading factor and wind speed similar to the
original \citet{Springel-03b} recipe.  Such a ``constant wind" model
gives roughly the correct global stellar mass density, but does not
fare as well at reproducing GSMFs~\citep[e.g.][]{Oppenheimer-10,Dave-11a}.
EAGLE, in contrast, does not decouple winds, but rather assumes
that ISM gas particles accumulate SN energy until they can be
instantaneously heated to a high temperature, typically $\gg 10^7$K
where cooling times are long and hence the thermal pressure can
drive an outflow.  While hydrodynamics remain turned on, cooling
is effectively turned off as the particle accumulates SN energy,
which occurs for tens of dynamical times, and hence this model is
also phenomenological.  In this way, the approach has similarities
to the blast wave cooling shutoff model of e.g.  \citet{Governato-07}
and \citet{Stinson-13}.  Nonetheless by tuning
this presciption, it is possible to nicely reproduce the $z=0$
stellar mass function, with a key aspect being that the given
amount of supernova energy from star formation is preferentially
injected into higher density gas~\citep{Crain-15}.  While
differing in approach, all these simulations have claimed to be in
broad agreement with observations of the GSMF.

In Figure~\ref{fig:mfsim} we compare the GSMF of the \mufasa\
fiducial $50\hmpc$ run to the Illustris simulation~\citep[orange dashed
line;]{Vogelsberger-14}, the EAGLE reference model~\citep[dark blue
dashed line;]{McAlpine-15}, and the MassiveBlack-II simulation~\citep[magenta
dashed line;][]{Khandai-15} at $z=0,1,2$ and 4.  The first two are
reproduced from \citet{Somerville-15}, while the latter data was kindly
provided by A. Tenneti and T. DiMatteo.  We omit error bars that
were shown in Figure~\ref{fig:mfres} to better see the differences,
but we include the same observations as were shown in that figure
for reference.

At $z=0$, the \mufasa\ and EAGLE simulations both do an excellent
job of reproducing the observed GSMF, with \mufasa\ doing somewhat
better at capturing the sharp knee of the mass function.  \citet{Crain-15}
describes the process of tuning EAGLE's input physics in order to
obtain this agreement, though it is worth noting that their mass
function mostly emerged rather naturally, and much of the tuning
was done to match galaxy sizes.  In our case, we mostly constrained
our input physics using higher-resolution simulations or analytic
models, but we still had to mildly tune some parameters such as the
outflow velocity.  Meanwhile, Illustris clearly overproduces galaxies
at both low and high masses, and only agrees well around $M^\star$
(i.e.  the knee).  MassiveBlack-II shows the characteristic steep
mass function that their ``constant wind" feedback model is known
to produce, which greatly overpredicts small galaxies and underpredicts
the GSMF around $M^\star$~\citep[see also][]{Dave-11a}.

At $z=1$, the discrepancies between simulations and data starts to
grow, particularly below $M^\star$.  All the simulations overpredict
the low-mass end to varying levels.  Illustris and MB-II are quite
high, and EAGLE is mildly high compared to observations.  \mufasa\
also overpredicts the low-mass end, but to a lesser extent than any
other model, making it clearly the best fit to the intermediate-redshift
GSMF.

At $z=2$, the discrepancies seen at $z=1$ for low mass galaxies
persist; generally, the simulations are signficantly higher than
observations here, with \mufasa\ providing the closest match.  At
the massive end, Illustris and MB-II nicely reproduce the turnover,
but EAGLE cuts off the GSMF at too low a mass.  \mufasa\ also
truncates the GSMF slightly early, albeit less so than EAGLE.  Part
of this may owe to the smaller $50\hmpc$ box size of \mufasa,
relative to the other simulations which have $\sim 70-100\hmpc$ box
sizes.

At $z=4$, Illustris and EAGLE do somewhat better than \mufasa, as
our fiducial \mufasa\ run produces too low a GSMF.  Our $25\hmpc$
volume which has comparable resolution to those runs, as well as
our $12.5\hmpc$ run, show better agreement as seen in
Figure~\ref{fig:mfres}.  As mentioned previously, the deficit in
the $50\hmpc$ run may owe to the lack of resolution to start galaxy
growth early enough, combined with the lack of volume required to
include growth of the largest halos.  Nonetheless, \mufasa\ is still
within $\sim 1\sigma$ of cosmic variance, so it is possible that
the chosen initial conditions unluckily do not produce enough early
massive galaxies.

While the \mufasa\ runs presented here lack a dynamic range comparable
to these larger simulations, they do offer several modeling advantages.
Compared to MassiveBlack-II, \mufasa\ employs a hydro solver that
has been shown to yield better results for key test
problems~\citep{Hopkins-15a}, and additionally our outflow scalings
better reproduce the observed GSMFs at the low-mass end.  Compared
to Illustris, \mufasa\ reproduces observed GSMFs significantly more
closely over most of cosmic time.  Generally, EAGLE and \mufasa\
yield comparably good mass functions, with \mufasa\ slightly better
from Cosmic Noon until today, and EAGLE slightly better at $z=4$.
\mufasa\ has some disadvantages as well.  Besides less dynamic
range, we also do not yet explicitly track central black hole growth
and feedback as the other simulations do (this is in progress; see
e.g. \citealt{Angles-16}).  Nevertheless, compared to recent
state-of-the-art galaxy formation simulations, \mufasa\ generally
is as good or better at reproducing the observed stellar mass buildup
in galaxies across the majority of cosmic time as traced by the
GSMF.

\section{Galaxy Growth Rates}\label{sec:sfr}

\subsection{Global star formation rate density}

\begin{figure}
  \centering
  \subfloat{\includegraphics[width=0.5\textwidth]{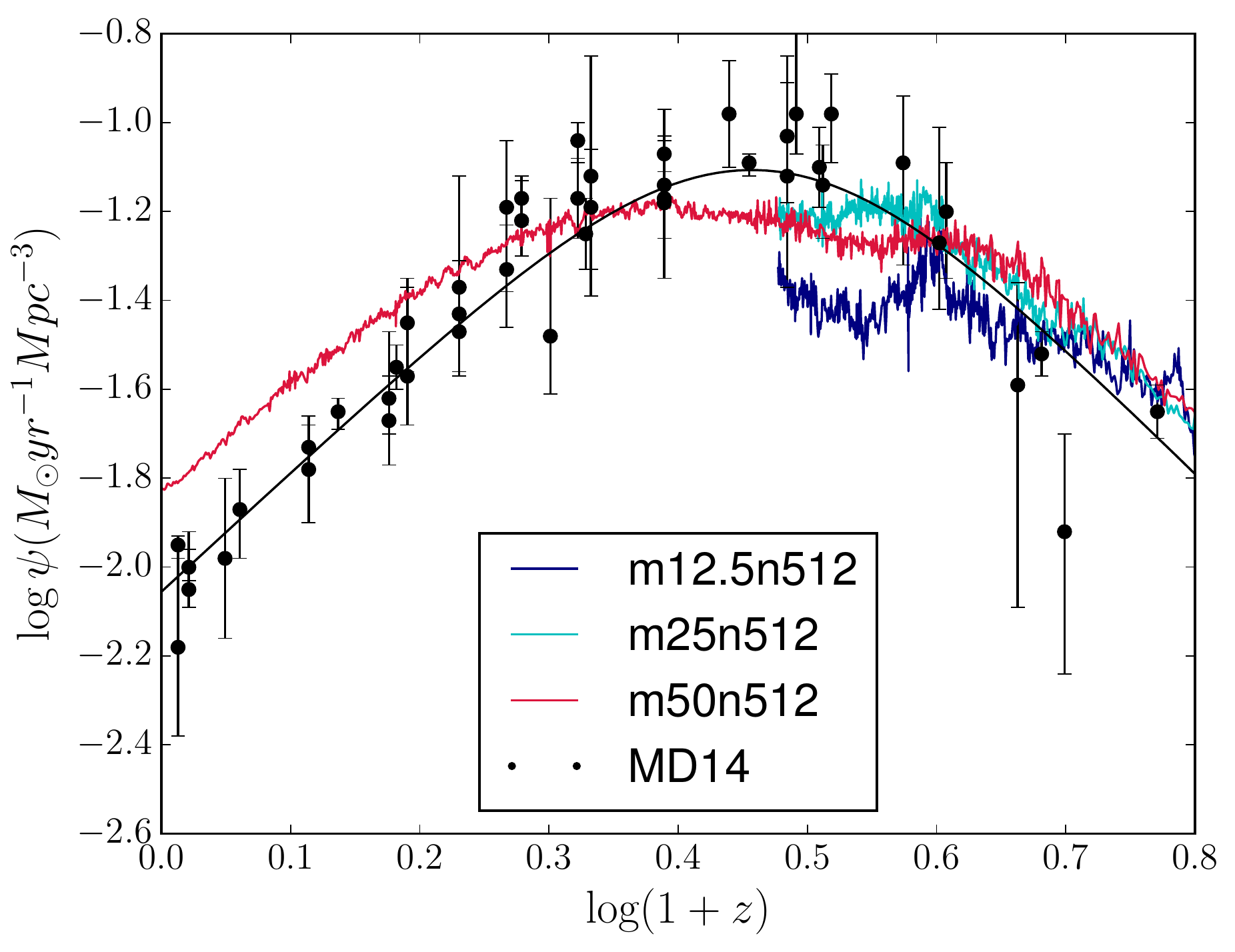}}
  \caption{ Cosmic star formation rate density evolution as a function of $\log(1+z)$, in 
our $50\hmpc$ (red), $25\hmpc$ (cyan), and $12.5\hmpc$ (dark blue) \mufasa\
simulations.  A fit to observations from \citet{Madau-14} is shown as the black line in each panel, and the individual data points from their compilation are also represented.
  	  }
\label{fig:madau}
\end{figure}

The growth of galaxy stellar mass is reflected in the star formation
rates of galaxies across cosmic time.  A key barometer of this is
the evolution of the cosmic star formation rate density (SFRD),
commonly known as a Lilly-Madau plot.  We note that the cosmic SFRD
is not an ideal test of models, because it is an integrated quantity
that can be sensitive to observational integration limits, and
individual simulations only cover a limited range of the total
cosmic star formation.  Hence discrepancies compared to observations
may reflect limitations of volume or resolution rather than intrinsic
model failings.  Nonetheless, the SFRD evolution is a oft-used
benchmark and highlights some important points about the characteristics
of galaxy growth.

Figure~\ref{fig:madau} shows the cosmic star formation rate density
($\psi$) evolution as a function of $\log(1+z)$.  Our three volumes
($50, 25, 12.5\hmpc$) are shown as the crimson, cyan, and navy blue
lines respectively.  Here, we have plotted the SFR directly output
during the simulation run as the stellar mass formed per unit time,
which displays small-timescale fluctuations owing to the stochastic
nature of our star formation algorithm; this employs no selection
criterion, and even includes star formation in unresolved galaxies.
For comparison, data from a recent compilation by \citet{Madau-14}
is shown in black, including a line for their global best-fit
relation along with the individual compiled data points; these have
been divided by 1.7 in order to correct from their assumed Salpeter
IMF to a Chabrier IMF.

The fiducial $50\hmpc$ simulation qualitatively follows the observed
trend faithfully: There is a rapid rise of $\psi$ from $z\sim
8\rightarrow 3$, then it is relatively constant until $z\sim 1.5$,
and then it falls rapidly towards the present day.  The evolution
since $z\sim 1.5$ generally follows the power-law expectation from
cosmological accretion, as has been noted by many previous
works~\citep[e.g.][]{Somerville-15}.  In detail, \mufasa\ underpredicts
the global SFR during Cosmic Noon (which we will discuss in
\S\ref{sec:ssfr}), and overpredicts the SFR at lower redshifts
($z\la 1.5$) by about 50\%.  This is particularly curious in light
of the good agreement with the evolution of the GSMF.  It is also
opposite to the discrepancy versus the recent models such as
Illustris, EAGLE, and the SAM of \citet{Henriques-15}, which tend
to {\it underpredict} the SFRD at low-$z$ while still generally
agreeing with GSMF evolution.  The SFRD peak at $z\sim 1-3$ is also
broader in \mufasa\ than in the data, which likewise mimics the
trend in other current models.  It is not obvious how one reconciles
all these models and observations; in part it may reflect systematic
uncertainties in star formation rate measures, or suggest that a
more careful comparison should be done to account for observational
selection effects.  We will break this down versus $M_*$ in
\S\ref{sec:ssfr} to gain more insights into the differences.

The $25\hmpc$ run matches the $50\hmpc$ quite well over the redshift
range probed.  This is a bit of a coincidence, as we have seen in
the GSMF that the low-mass end is slight overproduced in this run,
which appears to compensate for the lack of high-mass star-forming
galaxies owing to the smaller volume.  For the $12.5\hmpc$ run,
this coincidental balance fails, as the lack of high-mass star
forming galaxies dominates over the mild excess production at low
masses.  As a result, $\psi$ begins to lie significantly below that
of the larger volumes at $z\la 4$.  There is also significant
stochasticity over long timescales at $z\la 3$ (in addition to the
short stochasticity) because the global star formation is dominated
by a handful of the most massive galaxies whose star formation rates
can vary substantially.  Overall, the small-volume results are seen
to be not cosmologically representative at $z\la 4$ in terms of
large galaxy growth, but the evolution of dwarf galaxies should
still be robust.  At $z\ga 4$, all our volumes' SFRD values agree
with each other, and are within the uncertainties of the \citet{Madau-14}
data.

In summary, the SFRD evolution from $z=6\rightarrow 0$ is broadly
reproduced in our \mufasa\ runs.  This is not surprising given the
fact that the GSMF matches reasonably well at all redshifts $z\la
6$.  The deviations at $z\la 4$ in the $12.5\hmpc$ volume highlight
incompleteness at the massive end owing to its small volume; 
deviations at lower redshifts will be discussed \S\ref{sec:ssfr}.

\subsection{Star formation downsizing}\label{sec:sfh}

\begin{figure}
  \centering
  \subfloat{\includegraphics[width=0.5\textwidth]{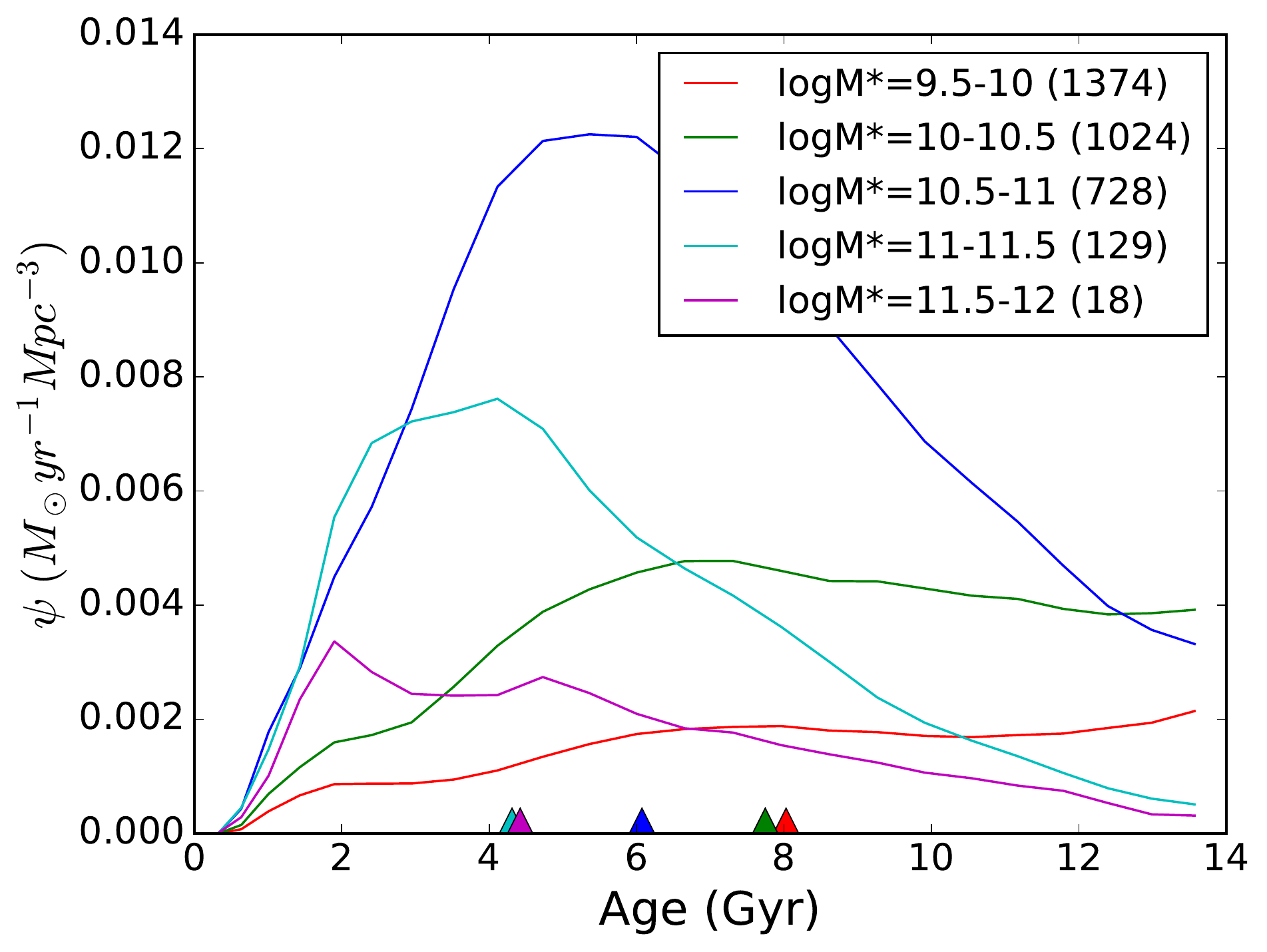}}
  \caption{ Cosmic star formation rate density evolution as a function of time in our fiducial $50\hmpc$ \mufasa\ run split into bins of final ($z=0$) galaxy stellar mass as indicated, with the number of $z=0$ galaxies shown in parantheses.  Colour-coded upward arrows along the bottom of plot indicate the time at which half the stars have formed within those galaxies.  Downsizing is evident, as more massive galaxies have an earlier median star formation time.}
\label{fig:sfhist_ms}
\end{figure}

A key characteristic of the galaxy population is known as ``downsizing",
first coined by \citet{Cowie-96}, in which star formation shifts
from more massive galaxies to lower-mass galaxies over time.  This
results in more massive galaxies having older stellar populations,
which has been dubbed ``archaeological downsizing"~\citep{Thomas-05}.
While downsizing has been referred to as anti-hierarchical, the
qualitative trend is a natural outcome of hierarchical structure
formation, since the largest galaxies arise in the largest density
perturbations which are able to collapse and start forming stars
first~\citep{Neistein-06}.  Nonetheless, quantitatively reproducing
the observed growth rates of galaxies as a function of mass has
proved challenging for models.

Since the fiducial \mufasa\ simulation well reproduces the observed
stellar mass function covering over 90\% of cosmic star formation,
it provides a quantitative picture of downsizing as a function of
galaxy mass.  Figure~\ref{fig:sfhist_ms} illustrates this.  Here,
we show the star formation history of galaxies subdivided into five
bins of $z=0$ stellar mass, from $10^{9.5}M_\odot$ to $10^{12}M_\odot$.
The time at which half the present-day stellar mass was formed is
indicated by the colour-coded upward arrows along the bottom of the
plot, for each mass bin.  Note that this is computed by examining
the formation times of the {\it remaining} stellar mass at $z=0$,
which is different than the star formation rate in those galaxies
at any particular epoch, owing to the effects of stellar mass loss;
in general, the median star formation rate will be shifted to
slightly earlier times.  This also makes no distinction between
stellar mass that was formed in the main progenitor versus accreted
satellite galaxies.  In this sense, our depiction here is most
closely related to archaeological downsizing.

Downsizing is clearly evident here -- higher mass galaxies have
earlier formation epochs.  Galaxies with $M_*>10^{10.5}M_\odot$,
which are predominantly quenched today, have median stellar formation times
around $4$~Gyr, or close to $z\sim 2$.  Galaxies with $M_*\la
10^{10.5}M_\odot$ in contrast have formation times of $\sim 8$~Gyr,
i.e. $z\sim 0.6$.  Galaxies around the knee of the mass function
today dominate the global stellar mass, and have a median formation
time of around 6~Gyr, or $z\sim 1$.  Hence our simulations naturally
predict that more massive galaxies will host older stellar populations.

The smallest galaxies have a global star formation rate that steadily
increases with time all the way to the present day, which is
qualitatively different than galaxies that end up around the knee
of the GSMF.  Such star formation histories in dwarfs are qualitatively
what is required to solve the so-called dwarf galaxy conundrum, in
which current models generically predict dwarf galaxy star formation
histories that peak at too early epochs~\citep{White-15}.  We will
examine this so-called dwarf galaxy conundrum using \mufasa\ in
more detail in future work.

\subsection{Stellar mass density evolution}\label{sec:rhostar}

\begin{figure}
  \centering
  \subfloat{\includegraphics[width=0.5\textwidth]{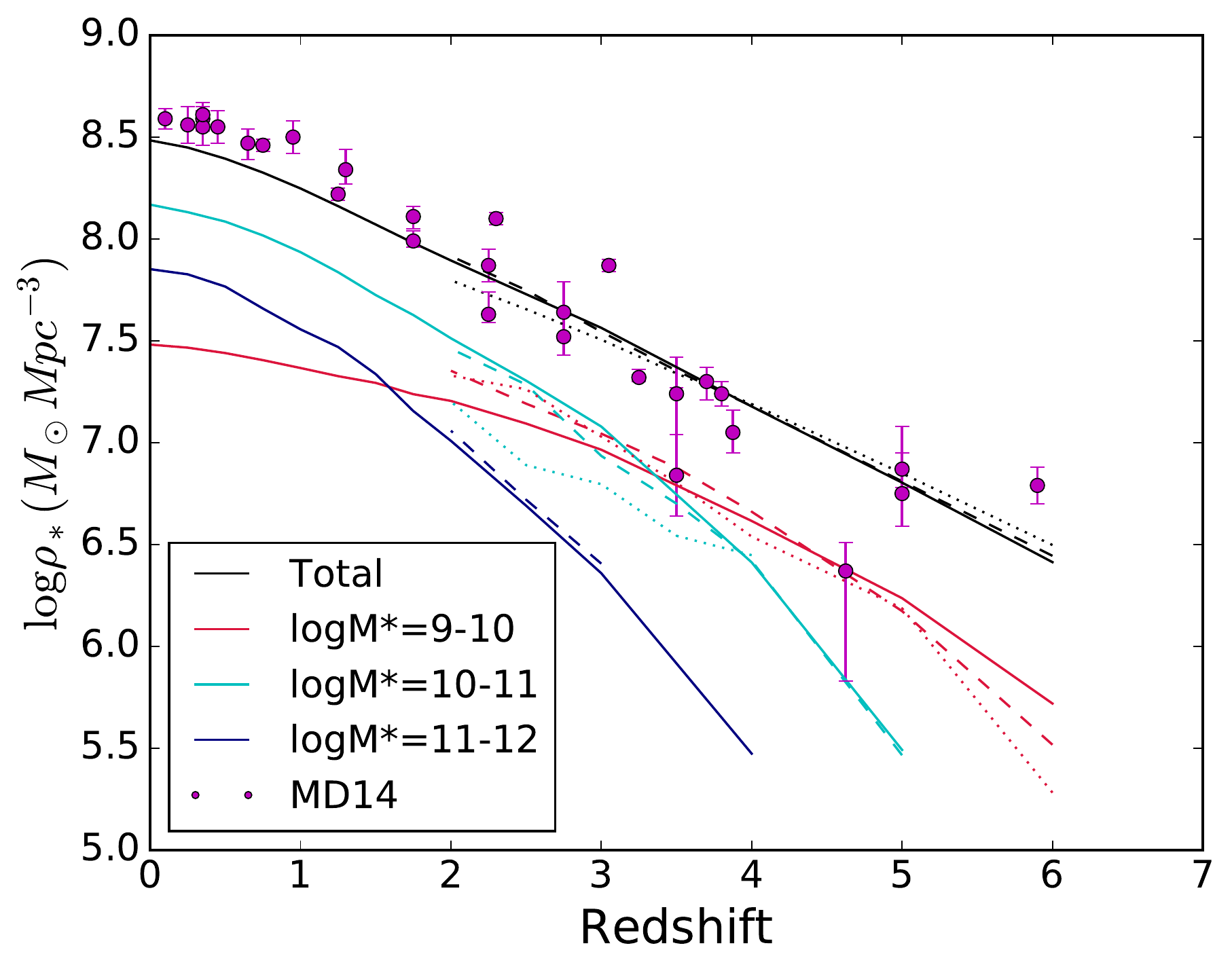}}
  \caption{ Cosmic stellar mass density evolution as a function of time in our \mufasa\ runs (black lines).  Solid lines show our $50\hmpc$ run down to $z=0$, while the dashed and dotted lines show the $25\hmpc$ and $12.5\hmpc$ volumes respectively.  Observational data is shown as compiled by \citet{Madau-14} (various sources described in text).  We further subdivide the global stellar mass into bins of stellar mass (at that epoch) as indicated.  The overall stellar mass density is in good agreement with data, and for most of cosmic time is dominated by galaxies with $M_*=10^{10-11}M_\odot$.}
\label{fig:rhostar}
\end{figure}

A complementary view of galaxy growth is provided by the evolution
of the cosmic stellar mass density.  In principle, this should be
equivalent to an integral of the Lilly-Madau plot, accounting for
stellar evolution.  Some discrepancies between these two approaches
have been claimed in the past~\citep[e.g.][]{Wilkins-08,Dave-08},
but these have been contradicted by more recent
data~\citep{Reddy-12,Madau-14}; the issue remains unsettled.

Figure~\ref{fig:rhostar} shows the evolution of the cosmic stellar
mass density in our three simulations; solid, dashed, and dotted
lines correspond to our $50, 25$, and $12.5\hmpc$ runs respectively.
We also show the breakdown in three ranges of stellar mass, the
dwarf ($10^{9-10}M_\odot$), $M^\star$ ($10^{10-11}M_\odot$), and
massive ($10^{11-12}M_\odot$) galaxy regimes, colour-coded as
indicated in the legend.  Finally, the cyan points show some recent
(post-2010) data from the compilation of observations by \citet{Madau-14},
namely \citet{Moustakas-13} at low redshifts, \citet{Ilbert-13},
\citet{Muzzin-13}, and \citet{Reddy-12} out to $z\sim 3$, and
\citet{Gonzalez-11}, \citet{Lee-12}, and \citet{Caputi-11} to higher
redshifts.

The total stellar mass density evolution is generally in good
agreement with observations.  At $z\la 1$ the $50\hmpc$ box falls
slightly below the data, potentially owing to observed galaxies
extending to masses below the resolution limit of this simulation.
The discrepancy is never more than 30\%, nonetheless it is interesting
that this contrasts with the predictions for the Lilly-Madau plot,
which show an {\it overprediction} of the observed SFRD at these redshifts.  This
hints at some mismatch between the observed global growth of stars
versus the observed global star formation rates.  At high redshifts
the agreement is very good, although the data shows more scatter
there.

Breaking $\rho_*(z)$ down by mass, at early epochs the total mass
density is dominated by relatively small galaxies, since larger
galaxies have not yet formed.  By $z\sim 3-4$, galaxies above
$10^{10}M_\odot$ begin to dominate the mass density, which continues
all the way until the present day.  Larger galaxies do not dominate
at early times since few have formed, and do not dominate at late
times since their growth is suppressed by quenching.  Examining the
various box sizes by comparing lines of the same colour, we see
that there is generally good convergence even when broken down by
mass, though there are variations of up to about 50\%.

Overall, \mufasa\ simulations do a good job of reproducing the
cosmic stellar mass density evolution, and the variations between
the different resolution boxes is not large.  Together with the
good match to the Lilly-Madau plot, this demonstrates that \mufasa\
provides a viable suite of models with which to examine global
stellar mass growth.

\subsection{Specific star formation rates}\label{sec:ssfr}

\begin{figure*}
  \centering
  \subfloat{\includegraphics[width=0.5\textwidth]{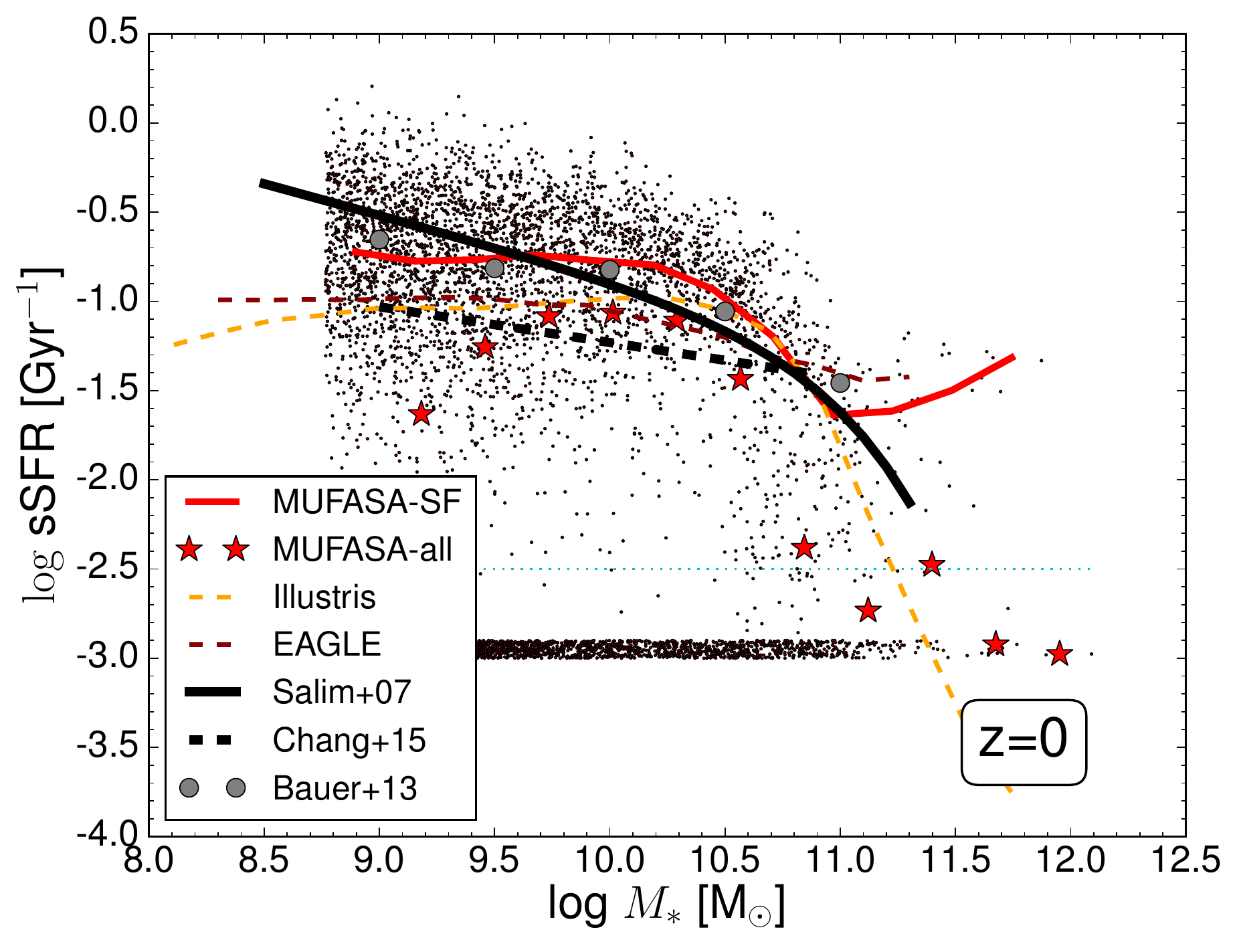}}
  \subfloat{\includegraphics[width=0.5\textwidth]{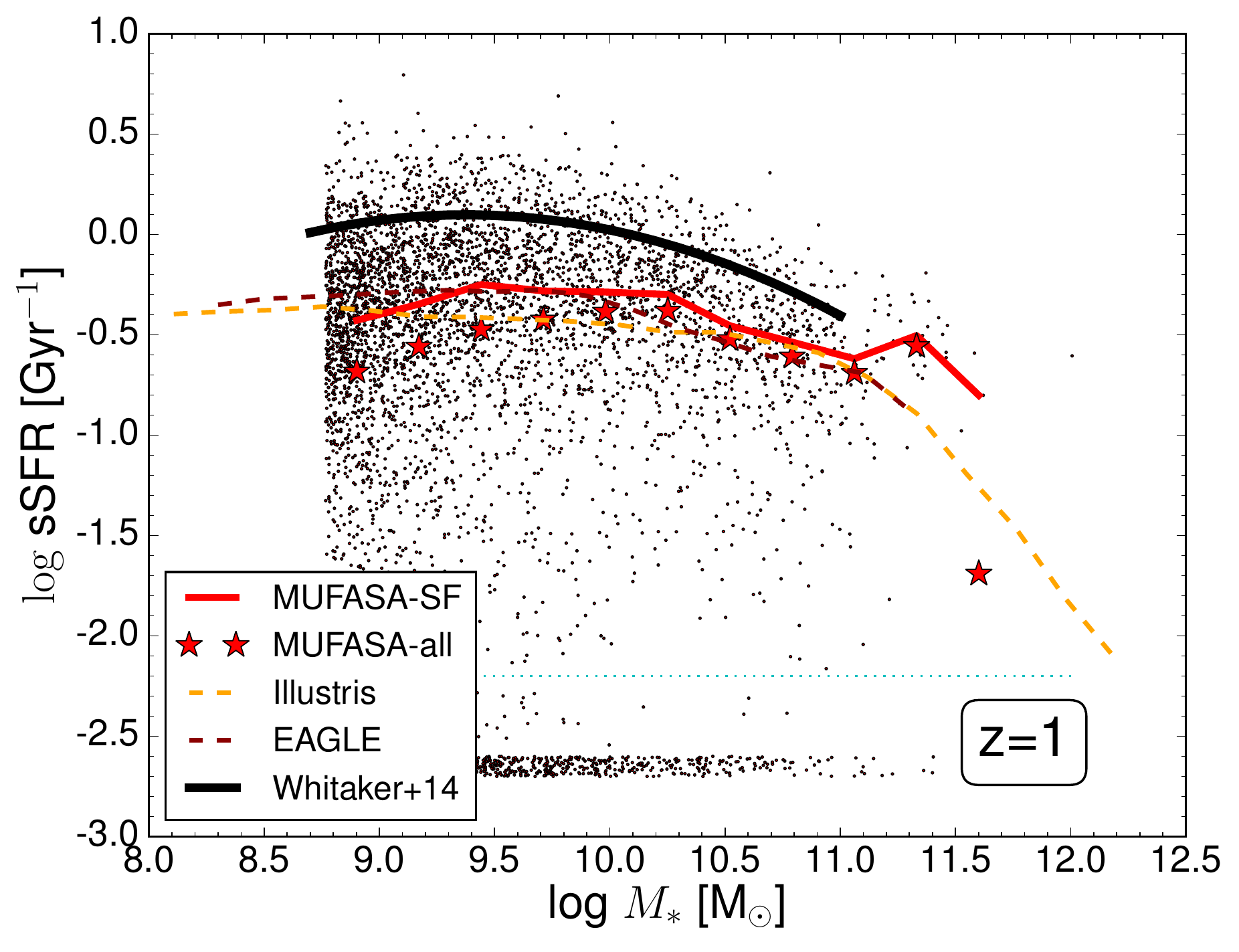}}
  \hfill
  \vskip-0.2in
  \subfloat{\includegraphics[width=0.5\textwidth]{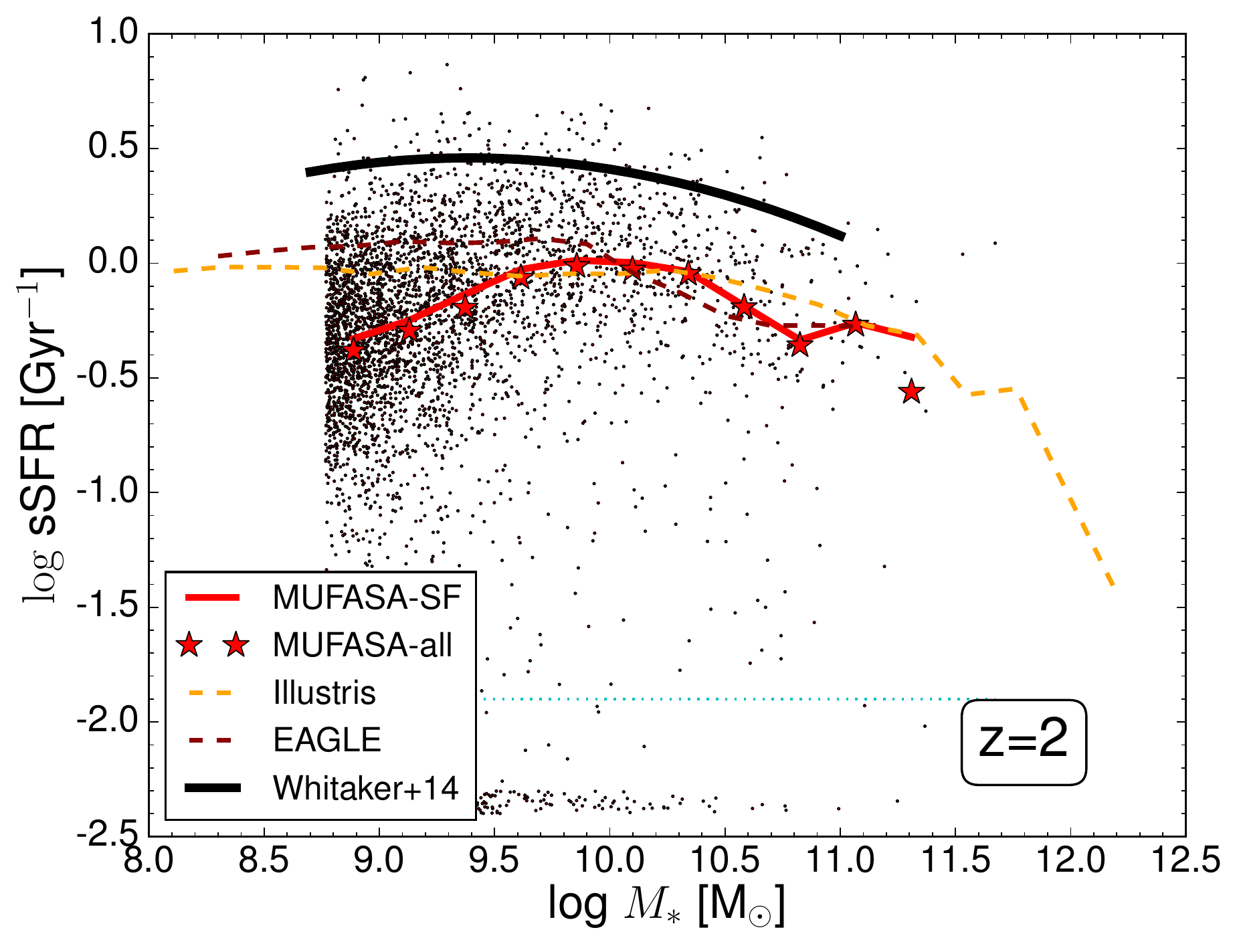}}
  \subfloat{\includegraphics[width=0.5\textwidth]{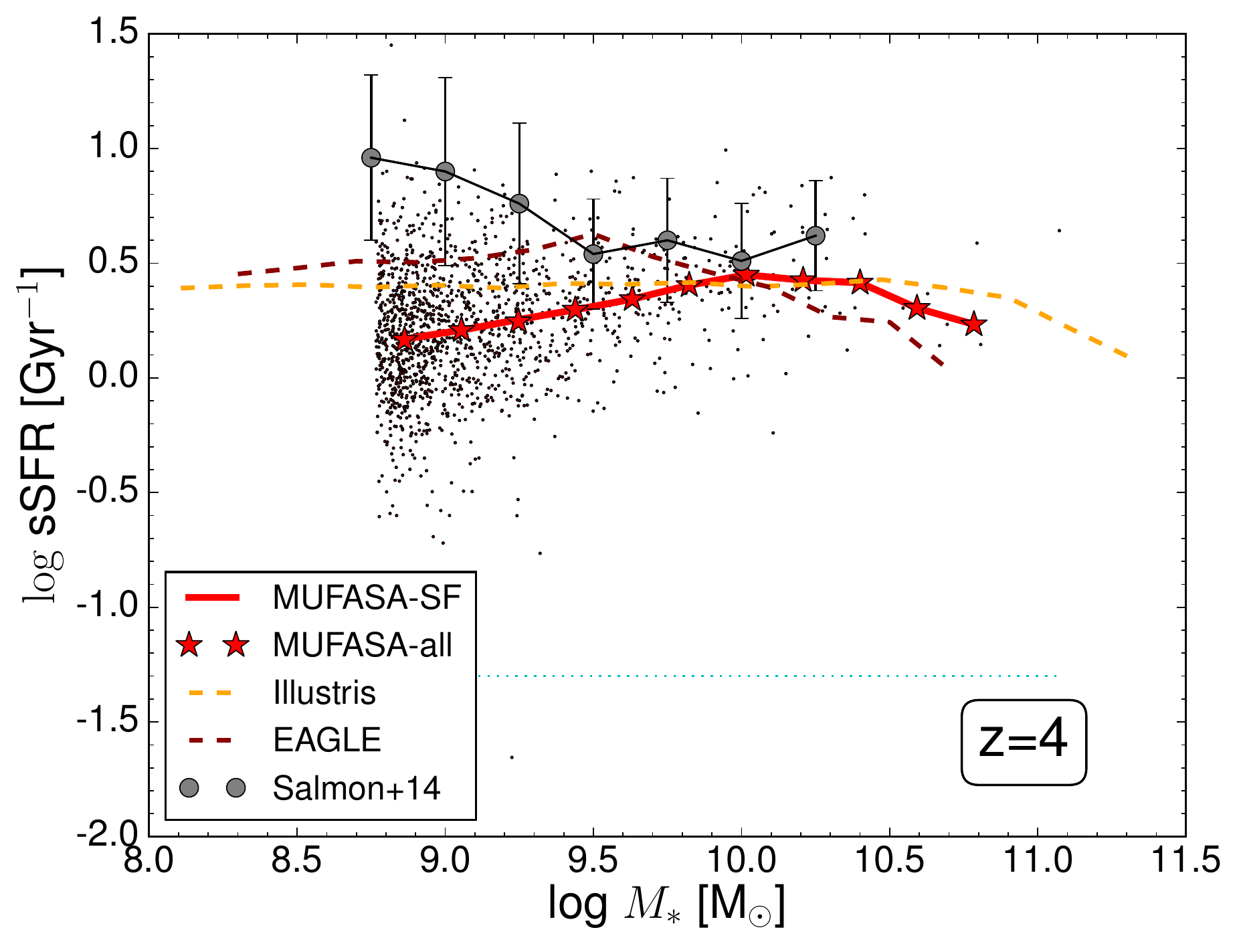}}
  \hfill
  \vskip-0.1in
  \caption{Specific star formation rate vs. stellar mass fo galaxies at $z=0,1,2,4$ in our fiducial $50\hmpc$ \mufasa\ simulation.  Points show individual galaxies, with galaxies having very low sSFR plotted across the bottom (with artificial scatter added for visibility).  Thick red lines show a running mean value for all star-forming galaxies, i.e. galaxies above the horizontal dotted cyan line near the bottom.  The stars show a running {\it median} sSFR, including non-starforming galaxies.  Observations are shown as follows:  At $z=0$, the Schechter function fit from \citet[black solid][]{Salim-07}, the fit from \citet[black dashed][]{Chang-15}, and data from \citet[grey points][]{Bauer-13}; at $z=1,2$ we show the polynomial fit to data from \citet{Whitaker-14}; while at $z=4$ we show data from \citet{Salmon-15}.  Finally, the dashed orange and dark red lines show results from Illustris and EAGLE, respectively.  The predicted sSFR is in good agreement at $z=0,4$, but is systematically low at $z=1,2$, following the trend seen in previous galaxy formation models.}
\label{fig:ssfr}
\end{figure*}

\begin{figure*}
  \centering
  \subfloat{\includegraphics[width=0.5\textwidth]{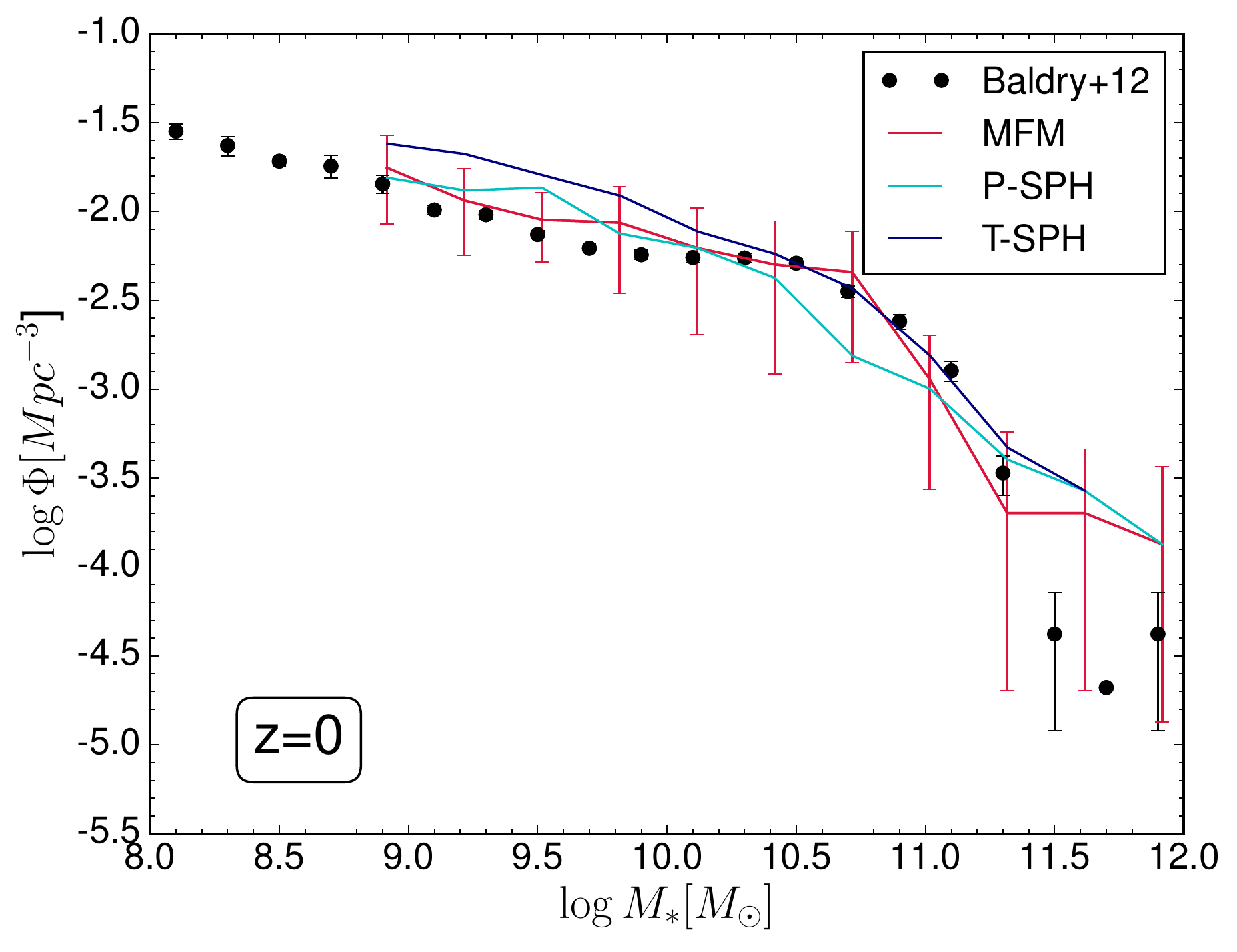}}
  \subfloat{\includegraphics[width=0.5\textwidth]{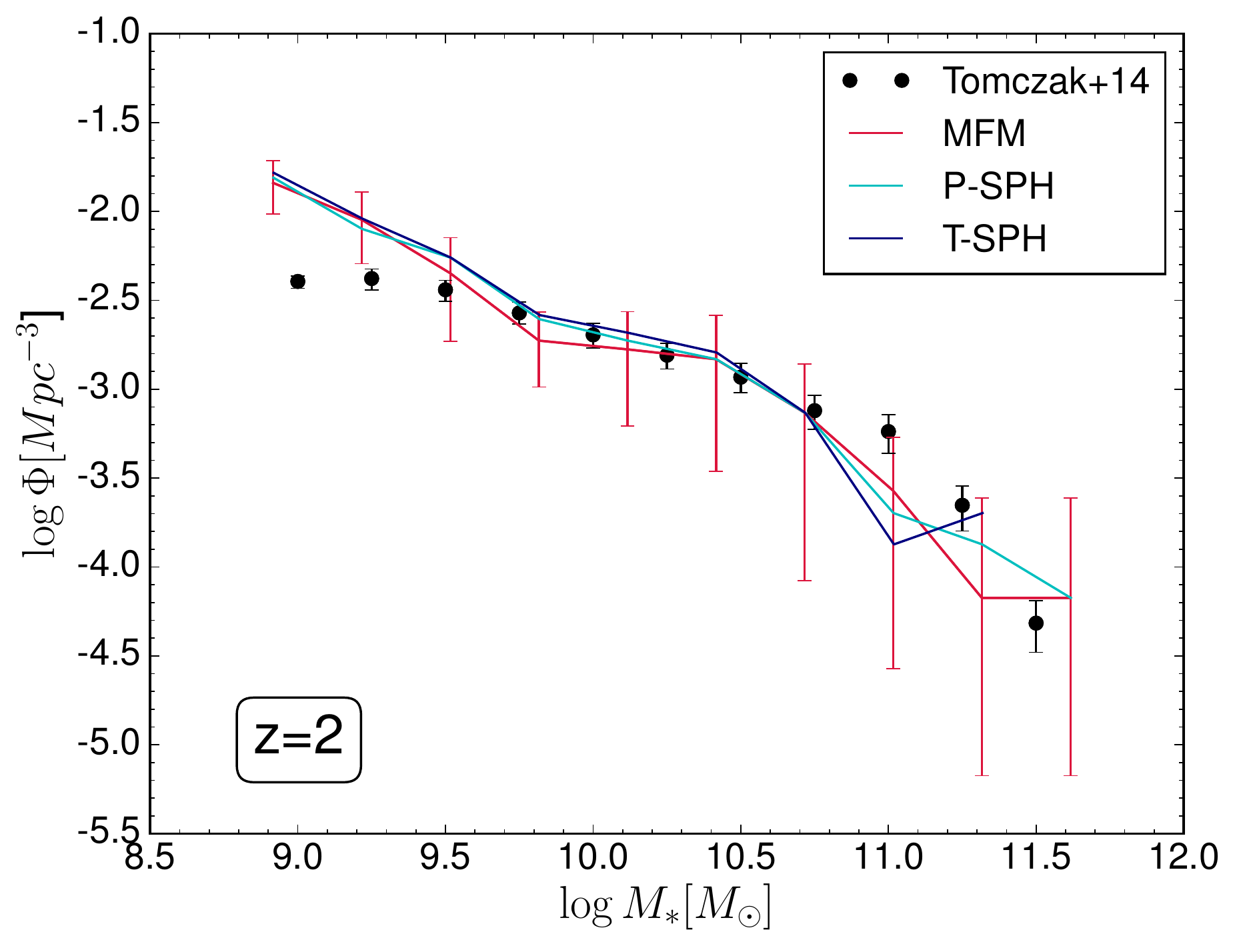}}
  \hfill
  \vskip-0.1in
  \caption{GSMFs at $z=0,2$ in $25\hmpc$, $2\times 256^3$ particle simulations run with three hydrodynamic solvers within {\sc Gizmo}: our fiducial Meshless Finite Mass (MFM; crimson), modern Pressure-SPH (P-SPH; cyan), and traditional SPH (T-SPH; navy).  All are started with identical initial conditions, with identical feedback physics.  Cosmic variance error bars are only shown for the MFM case; the others are comparable.  MFM and T-SPH tend to produce larger galaxies than P-SPH, more prominently so at later epochs, while MFM produces a flatter low-mass slope.}
\label{fig:mfhydro}
\end{figure*}

A key test of how galaxies grow is the evolution of the star formation
rate--stellar mass relation, often called the star-forming galaxy
``main sequence"~\citep{Noeske-07}.  Now measured out to $z\sim 6$
and beyond, its evolution has proved to be challenging for models
to reproduce particularly during Cosmic Noon ($z\sim1-3$), with
both SAMs and simulations typically predicting specific star formation
rates (sSFRs) at $z\sim 2$ that are too low by $\sim\times
2-3$~\citep{Daddi-07,Dave-08,Sparre-15,Somerville-15}.  Many of
these same models, however, also tend to overpredict the GSMF during
Cosmic Noon by roughly the same factor, so it has been conjectured
that a model that matches the GSMF at $z\sim 1-3$ (and keeps the
SFRs the same) would also match the main sequence.  Since the
\mufasa\ simulations provide a reasonable fit to the GSMF at $z=1-3$,
it is interesting to examine whether they likewise matches the main
sequence at these epochs.

Figure~\ref{fig:ssfr} shows sSFR vs. $M_*$ at $z=0,1,2,4$ in our
fiducial $50\hmpc$ \mufasa\ simulation.  Along the bottom are
depicted galaxies with very low (typically zero) sSFR, with some
artificial spread to enhance visibility.  The red solid line shows
the running median sSFR as a function of $M_*$, for star-forming
galaxies defined as those above the cyan dashed line lying just
above the quenched galaxies.  The stars indicate a binned median
for all galaxies, including the quenched ones.  Observations are
shown in black:  At $z=0$, we show a Schechter-form fit from
\citet[solid black]{Salim-07} based data from the Sloan Digital Sky
Survey (SDSS) and the Galaxy Evolution Explorer (GALEX), as well
as H$\alpha$-based star formation rate determinations from the
Galaxy Mass and Assembly (GAMA) survey by \citet[gray points]{Bauer-13}.
We also show a fit to the recent SDSS+Wide Infrared Survey Explorer
(WISE) determination by \citet[dashed black]{Chang-15}.  At $z=1,2$
we show data from 3D-HST with {\it Spitzer} 24$\mu$m information
by \citet{Whitaker-14}, and at $z=4$ from CANDELS by \citet{Salmon-15}.
Finally, for comparison we show results from the Illustris (orange
dashed) and EAGLE (dark red dashed) simulations.

At $z=0$, \mufasa\ star-forming galaxies show a flat low-mass end
and a turn-down at $M_*\ga 10^{10.5}M_\odot$ as the general population
starts to be quenched (but is still regarded as star-forming by our
cut).  The predictions are in very good agreement with recent
H$\alpha$-based SFR's from GAMA \citep[grey circles]{Bauer-13}, and
in reasonable agreement with the SDSS+GALEX data from \citet{Salim-07},
However, they are higher than the sSFR's inferred from full $0.4-22\mu$m
SED fitting including WISE data \citep{Chang-15}.  It is beyond the
scope of this paper to assess the various differences in observations,
but we note that even the present-day SFRs are somewhat controversial.

For comparison, we show results from Illustris (orange; $z=0$) and
EAGLE (dark red; $z=0.1$) simulations, which like \mufasa\ show a
flat sSFR($M_*$) at low masses~\citep[see also][]{Dave-11a}, but
with an amplitude that is lower by $\sim\times 1.5-2$.  These are
generally in better agreement with the \citet{Chang-15} data.  Hence
even simulations that nicely match the GSMF such as \mufasa\ and
EAGLE can have substantially different $z=0$ sSFR values for
star-forming galaxies.  As mentioned earlier, the global SFRD in
these other simulations tend to be lower than that in \mufasa, and
from the sSFR$(M_*)$ comparison this is likewise seen to be the
case for individual galaxies with $M_*\la 10^{10.7}M_\odot$.  There
are also a handful of massive \mufasa\ galaxies that are still
star-forming as well, which contribute non-trivially to the global
SFRD.  We will examine the massive galaxy population and its colours
in an upcoming paper.

The stars show the binned median values from \mufasa\ now including
the quenched galaxies.  Such galaxies dominate by number at low and
high masses -- the median values (stars) at $M_*\sim 10^{9}M_\odot$
and $M_*\ga 10^{11}M_\odot$ are actually in the quenched population.
At low masses these are typically stripped satellites, while at
high masses they are quenched centrals.  Observations generally do
not suggest quite such a dominant population of low-mass quenched
satellites, but such satellites are also difficult to observe, so
selection effects may be important.  We leave a more thorough
investigation of central vs. satellite populations for future work.

Moving to $z=1$, the \mufasa\ predictions clearly lie below the
data, typically by a factor of up to 2 relative to the \citet{Whitaker-14}
data across the resolved mass range.  Illustris and EAGLE are
comparable.  Indeed, such a discrepancy is generically seen among
models of all types~\citep{Somerville-15}.  The number of quenched
galaxies is markedly less, so that the medians are now generally
within the star-forming population.

At $z=2$, the agreement gets even worse -- the deficit of \mufasa\
versus the \citet{Whitaker-14} data is now $\times 3$ across most
of the mass range.  This is comparable to or even lower than the
predictions from Illustris~\citep{Sparre-15} and EAGLE~\citep{Schaye-15}.
There are very few quenched galaxies by this epoch, so one cannot
easily appeal to burstiness to explain the discrepancy -- besides
the fact that the observed tight main sequence scatter limits the
contribution of starbursts~\citep{Rodighiero-11}.

This discrepancy at $z\sim 2$ remains a major puzzle, and it starkly
contrast with the fact that \mufasa\ nicely reproduces the evolution
of the GSMF around this redshift range.  If the Universe is producing
stars at a rate $\sim 3$ times higher than \mufasa\ at $z\sim 2$,
and the GSMF agrees at $z\sim 2$, then where do all those stars end
up?  They don't appear to produce an excess in the observed GSMF
relative to \mufasa\ at $z\sim 1$; if anything, \mufasa\ lies
slightly {\it above} the observed GSMF at $z=1$.  Hence it seems
that the rate of stellar growth as tracked by the GSMF is not
straightforward to reconcile with that measured directly via star
formation rates.  It could be that the observed star formation rates
are simply too high owing to calibration issues, but the uncertainties
in UV+24$\mu$m star formation rates are generally thought to be
significantly smaller than the discrepancy.  As the conjecture in
\citet{Bastian-10} states, any problem in galaxy formation can be resolved
with a suitable choice of IMF, and in this case a top-heavy or
bottom-light IMF remains a viable if unpopular
solution~\citep{Wilkins-08,Dave-08}.

The discrepancy could also be solved by raising $M_*$ values at
$z\sim 2$ by $\sim\times 3$ via some unforeseen systematic errors.
Given that current GSMF data at this epoch is now based on deep
near-IR data that probes well beyond the 4000\AA\ break, it would
be quite surprising if the stellar mass estimates were so far off;
current methods for estimating $M_*$ typically do not differ
systematically by such large amounts~\citep{Mobasher-15}.  Furthermore,
this would shift the GSMF to quite high values, and would result
in their being more massive galaxies at $z=2$ than today.  Hence
this solution seems untenable.

Finally, at $z=4$ the dynamic range spanned by observations is
relatively small, but the disagreements versus the data are clearly
less, particularly for larger star-forming galaxies.  \mufasa\ shows
little trend with mass at this redshift, and perhaps even has a
positive slope at $M_*\la 10^{10}M_\odot$.  The \citet{Salmon-15}
data plotted shows a rise towards the lowest masses that at face
value is divergent from the model predictions, but incompleteness
in identifying low-SFR galaxies via Lyman break selection at these
redshifts may mitigate this disagreement.  The discrepancy in
sSFR($M_*$) versus models thus seems to be primarily restricted to
Cosmic Noon ($z\sim 1-3$), as has long been noted~\citep[e.g.][]{Dave-08}.

In summary, \mufasa\ shows similar trends to previous models in
matching observations at low and high redshifts, but falling low
throughout Cosmic Noon.  At $z=0$ there are substantial differences
versus recent simulations at the low-mass end, with \mufasa\ showing
higher levels of present-day star formation in star-forming dwarfs,
although quenched dwarfs dominate by number.  At intermediate
redshifts, \mufasa\ mimics the shape if not amplitude of sSFR($M_*$),
with a positive sSFR($M_*$) slope at low masses and a negative one
at high masses.  The high observed sSFR values across all masses
at $z\sim 1-3$ remain a persistent quandary for modern galaxy
formation models, even ones that match the observed GSMF evolution
through that epoch.

\begin{figure*}
  \centering
  \subfloat{\includegraphics[width=0.5\textwidth]{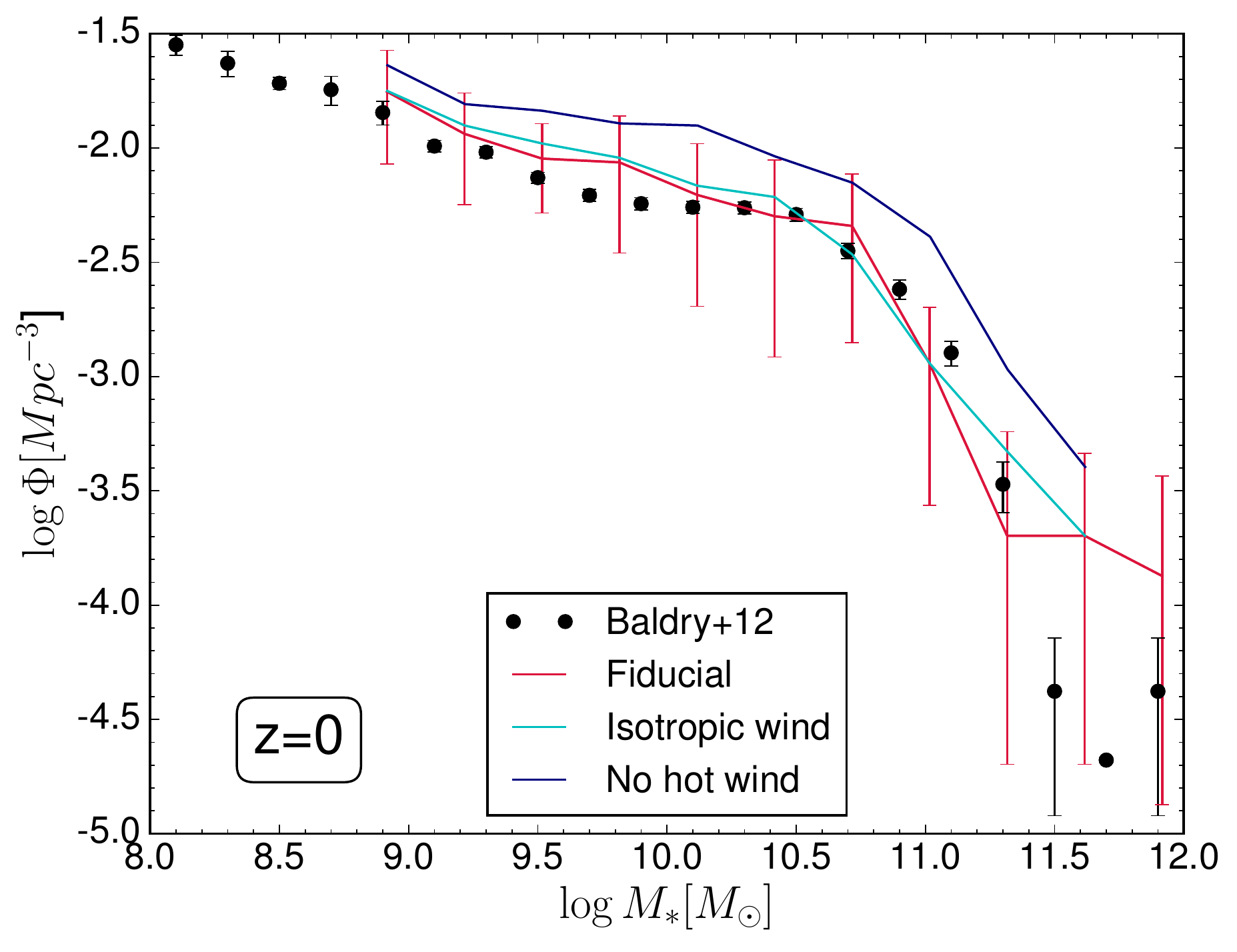}}
  \subfloat{\includegraphics[width=0.5\textwidth]{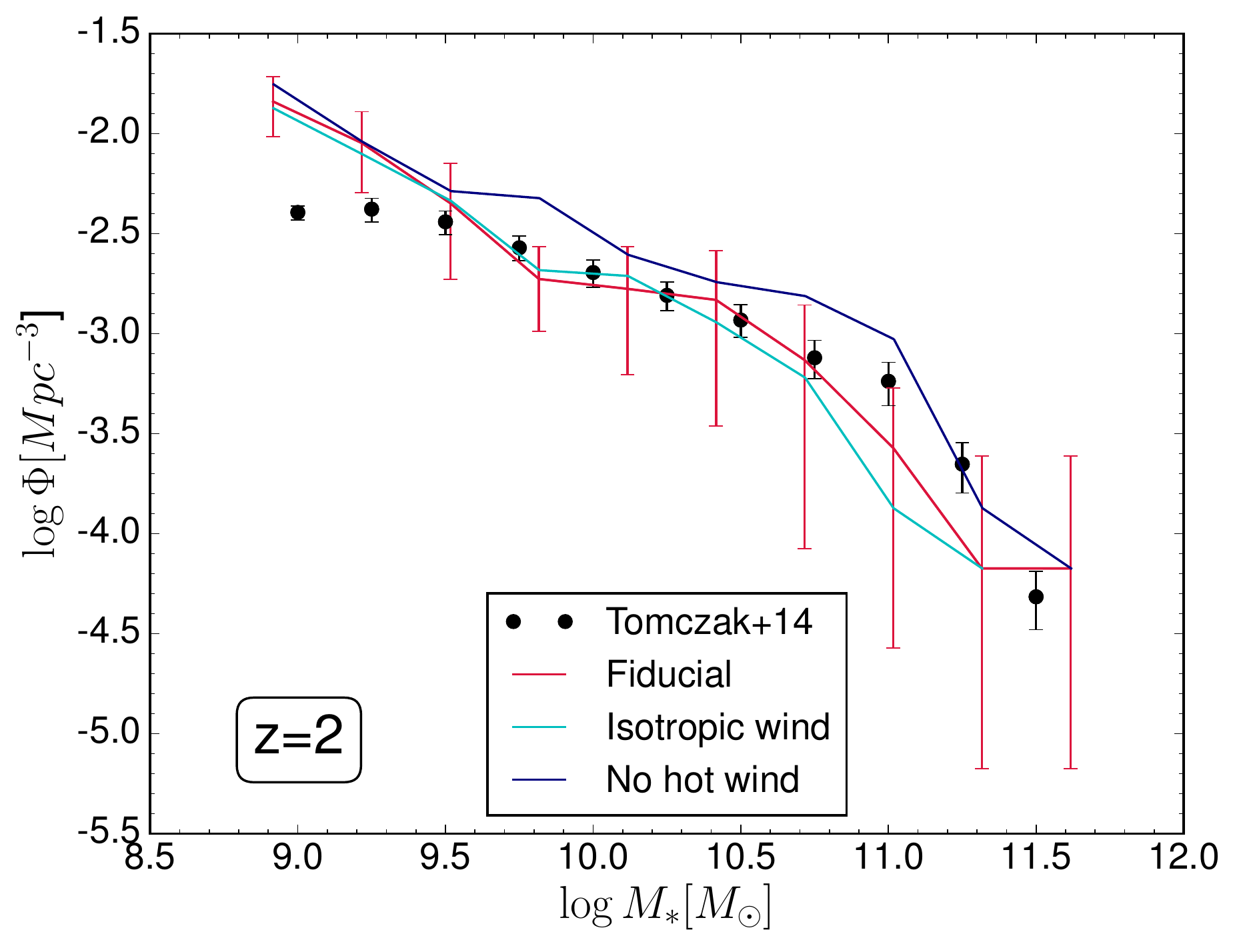}}
  \hfill
  \vskip-0.1in
  \caption{GSMFs at $z=0,2$ in $25\hmpc$, $2\times 256^3$ particle simulations runs comparing our fiducial feedback prescription (crimson) with one with purely cool winds (navy) and having isotropic wind ejection (cyan).  Cosmic variance error bars are only shown for the fiducial case; the others are comparable.  The inclusion of 30\% hot winds in the fiducial case significantly suppresses galaxy growth, moreso at later epochs, while ejecting winds isotropically does not make a substantial difference.}
\label{fig:mfwind}
\end{figure*}

\section{Code variations}\label{sec:codevar}

A major advantage of \gizmo\ is that it has the ability to trivially
switch between four hydrodynamic solvers while holding all other
input physics fixed.  Hence it provides an ideal platform for testing
the sensitivity of results such as the GSMF to hydrodynamic
methodology.  To contextualise the differences resulting from
numerical methodology, we explore how key input physics choices
change the GSMF.

\subsection{Sensitivity to hydrodynamics methodology}\label{sec:hydro}

\gizmo\ contains four hydrodynamics solvers: MFM, which is our
fiducial choice; meshless finite volume (MFV), which is a mass-advecting
moving mesh-like scheme; pressure-SPH~\citep[P-SPH;][]{Hopkins-13},
which is a pressure-energy formulation of SPH that improves surface
instability handling; and traditional SPH (T-SPH), which is a
density-energy formulation of SPH.  These are described in more
detail in \citet[Appendix
F]{Hopkins-15a}.  We note that T-SPH employs some new features in
{\sc Gizmo} relative to older SPH codes, such as an improved
artificial viscosity and artificial conduction; see \citet{Hopkins-15a}
for details.  Since MFV does not strictly conserve mass within fluid
elements (though it does so approximately), it is not obvious how
reliably it will function with the kinetic outflows, stellar mass
loss, etc., hence we will not consider this scheme here.  Instead,
we will focus on comparing the mass-conserving schemes MFM, P-SPH,
and T-SPH.

Figure~\ref{fig:mfhydro} shows the GSMF at $z=0$ (left) and $z=2$
(right) in runs with MFM, P-SPH, and T-SPH.  These runs have a box
size of $25\hmpc$ and $2\times 256^3$ particles, but in every other
way are identical to our fiducial $50\hmpc$ \mufasa\ run, excepting
the hydro solver which is only identical for the MFM case.  Observations
are shown as described in Figure~\ref{fig:mfres} for reference,
although the focus here is on comparing among the hydro solvers
rather than comparing to data.  Error bars are again computed as
cosmic variance over eight sub-octants within each simulation volume
(only shown for the MFM case; the others are similar but omitted
for clarity), and they are larger than in the full $50\hmpc$ \mufasa\
run owing to the smaller volume.

We first note that the $25\hmpc$ MFM run produces a GSMF that is
statistically identical to the $50\hmpc$ run.  This is not surprising,
as they have identical numerical resolution and input physics, but
it is reassuring that these simulations display good GSMF convergence
with volume, even though there are many fewer galaxies in the smaller
volume.  

Comparing amongst the various hydro solvers, at early epochs
($z=2$) the GSMFs are very similar.  In other words, kinetic outflows
are suppressing early galaxy formation in a manner that is essentially
independent of the hydro solver.  Indeed, the differences at $z\geq
3$ (not shown) are even smaller.

By $z=0$, the three hydro solvers diverge non-trivially in their
predictions.  As discussed in \citet{Oppenheimer-10}, the contribution
from wind recycling (i.e. the re-accretion of previously ejected
material) to stellar mass growth starts to be significant at $z\la
2$, and becomes dominant by $z=0$.  Our preliminarily interpretation
is thus that the differences between hydro solvers owe primarily
to differences in the way that galactic outflows interact with the
surrounding medium and rejoin the accretion flow.  This is consistent
with the idea that the main improvement in MFM over SPH is handling two-phase
instabilities and shocks, which are important mechanisms for how
winds interact with circum-galactic gas.

T-SPH tends to produce about $50\%$ more galaxies at any given
$M_*$ compared to P-SPH.  Meanwhile, both SPH codes produce a steeper
GSMF than MFM.  In \citet{Oppenheimer-10}, it was noted that the
flattening of the GSMF in the sub-$M^\star$ regime is achieved in
simulations by wind recycling.  If this is likewise the case in
these simulations (as we expect), then we can infer that winds
recycle more effectively in MFM than in SPH.

The detailed reasons for these variations are not entirely obvious.
Our preliminary interpretation for difference between P-SPH and
T-SPH is as follows.  In flows with large variations of internal
energy $u$ from particle-to-particle, P-SPH, because it kernel-averages
the pressure over all neighbors' $u$ as opposed to just using the
particle's $u$ times density, a single very high-$u$ particle can
raise the pressure of all its neighbors.  This means that all of
them together can do more PdV work than a single hot particle could.
While this is done in an energy-conserving way, it can effectively
enhance the ability of lone, hot particles to add substantial
pressure.  In our case, the thermal heating of the winds results
in such lone hot particles, which then increases the pressure in
CGM gas upon recoupling, moreso in P-SPH than T-SPH.  This can
then expand the CGM gas and lower the amount of wind re-accretion,
which results in a lower GSMF for P-SPH.  While we have not confirmed
this effect in detail, we believe this is a plausible interpretation
for the origin of the discrepancy between P-SPH and T-SPH.  Meanwhile,
MFM generally increases the amount of mixing, which increases the
amount of recycling relative SPH, and hence flattens the GSMF.

This highlights a key point that, although we have left all the
feedback physics and numerical parameters fixed in the code between
the three methods, there are inherent differences between the methods
that can interact with feedback in ways that depend on methodology.
For instance, \citet{Hopkins-15a} points out that the effective
resolution in discontinuities and shocks is higher in MFM versus
SPH owing to less smoothing, which could play a role in more
accurately modeling the hot wind interactions.  We also note that
the details of SPH scheme can matter, and differences can arise
when injecting energy using SPH owing to the manner in which energy
or entropy is smoothed among neighbors~\citep{Schaller-15}.  The
sensitivity to methodology presented here therefore should be
regarded as specific to both the hydro scheme and the feedback
implementation, and not a generic result.  Nonetheless, it gives
some idea as to the level of variations that can occur among modern
methods.

It is of course not immediately evident which hydro solver yields
a more correct answer.  Nonetheless, in idealised test cases, MFM
handles discontinuities and shear flows in a more realistic manner
than SPH~\citep{Hopkins-15a}, which is consistent with results in
more realistic circumstances~\citep{Wetzel-16}.  We expect that
issues of hydrodynamics will be more important in massive galaxies
that contain hot gaseous halos, where two-phase discontinuities are
stronger; hence our $25\hmpc$ volume is perhaps not ideal for these
tests.  For now, we preliminarily conclude that there are differences
of up to $\sim\times 2$ for $z\la 1$ GSMF predictions based purely
on which hydrodynamics solver is used, while at higher $z$
the predictions are less sensitive to this.  

\subsection{Sensitivity to outflow parameterisations}\label{sec:phys}

\mufasa\ includes new implementations of feedback prescriptions,
so it is interesting to investigate how some of these novel aspects
impact the predicted GSMFs.  Here we examine two aspects, namely
two-phase winds and the wind ejection direction, and determine how
large of an impact they make on the resulting GSMFs.

Figure~\ref{fig:mfwind} shows the GSMF at $z=0$ (left) and $z=2$
(right) in simulations where we turn off two-phase winds and eject
all winds cool (i.e. at $T\sim 10^4$K), shown in navy blue, while
the cyan curve shows the result of ejecting winds isotropically
rather than outwards in the {\bf v}$\times${\bf a} direction.  It
is clear that the choice of wind direction has a very minimal impact
on the resulting GSMF, though we might expect that it could have
signatures in the enrichment of the CGM.

In contrast, turning off the hot wind component has a large impact,
somewhat more so at later epochs; at $z\ga 4$ (not shown) the
differences are fairly small.  Evidently, the hot component of outflows 
adds significant preventive feedback, thereby suppressing galaxy
growth.  Number densities are lowered by $\sim\times 2-3$, with
a somewhat greater impact for more massive galaxies.
Ejecting some fraction of winds hot is therefore an important
aspect of how \mufasa\ obtains good agreement with the GSMF.

Besides indirectly impacting the GSMF, two-phase outflows are also
likely to significantly impact the thermal state of the CGM.  Hence
observations of quasar absorption lines around galaxies such as the
COS-Halos project~\citep{Tumlinson-13} can in principle provide
constraints on outflow temperatures.  Previous simulations that ejected
winds purely cold tended to underpredict the amount of absorption
from ions tracing warmer gas such as \ion{O}{vi}~\citep{Ford-16}.
In future work we will examine whether the new two-phase outflow
model produces better agreement.

In summary, the thermal state of the outflows adds an important
component of preventive feedback that helps suppress galaxy formation
in accord with observations.  The choice of wind direction, in
contrast, has minimal impact.  Zoom simulations that self-consistently
drive outflows via supernova heating can in principle directly
predict the temperature distribution of outflowing gas, and we hope
that future such efforts will aim to quantify predictions for the
thermal state of outflowing gas, in addition to the mass loss rate
and velocity.  We note that the differences in the GSMF owing to
two-phase winds exceed the differences owing to choice of hydrodynamics
methodology discussed in the last section; this is consistent with
previous investigations~\citep[e.g.][]{Schaye-10,Dave-11a,Hopkins-14}.

\section{Summary}\label{sec:summary}

We present a new set of cosmological hydrodynamic simulations called
\mufasa\ run using the meshless Godunov finite mass hydrodynamics
method in the new {\sc Gizmo} code~\citep{Hopkins-15a}.  We include
updated recipes for star formation and feedback, including molecular
hydrogen based star formation following~\citet{Krumholz-09}, two-phase
kinetic outflows with scalings based on the FIRE zoom simulations
of \citet{Muratov-15}, and halo mass based quenching with an evolving
quenching mass scale taken from analytic equilibrium model constraints
by \citet{Mitra-15}.  Our feedback prescriptions are still 
phenomenological, but the parameters are mostly taken from high-resolution
simulations or independently-constrained analytic models rather
than tuned by directly matching our simulations to observations.
We run simulations with $50\hmpc$, $25\hmpc$, and $12.5\hmpc$, the
large fiducial volume to $z=0$, and the latter two to $z=2$ to study
the high-redshift population in more detail, each having $2\times
512^3$ particles and sub-kpc resolution.  We test our models against
measures of galaxy growth from $z=6\rightarrow 0$, particularly the
galaxy stellar mass function, the global evolution of stellar mass
density and star formation rate, and the stellar mass--star formation
rate relation.  We briefly investigate how much choices of hydrodynamics
methodology and outflow modeling impact the predicted GSMF.

Our main results are summarised as follows:
\begin{itemize}

\item The fiducial $50\hmpc$ \mufasa\ simulation agrees very well
with the observed evolution of the GSMF from $z=6\rightarrow 0$,
usually matching data to $\sim 1\sigma$ in cosmic variance uncertainty
or better, and providing unprecedented agreement to this key barometer
for galaxy formation models.  The resolution convergence between
the three volumes at $z\geq 2$ is reasonable good, with a systematic
increase of $\times 2$ in galaxy number density over a factor of
64 in mass resolution.

\item Compared to recent hydrodynamic simulations Illustris, EAGLE,
and MassiveBlack-II, \mufasa\ reproduces the observed GSMF about
equally well or better at any given redshift.  Particularly, it has
been challenging for models to match the GSMF at $z\sim 1-3$, which
\mufasa\ does quite well.  This suggests that the cosmic growth
rate of stellar mass in galaxies can now be viably modeled by
cosmological hydrodynamic simulations.

\item The cosmic star formation rate density evolution is in general
agreement with observations, albeit too low during Cosmic Noon and
too high at $z\la 1.5$.  The cosmic stellar mass density growth
shows a strong trend of archaeological downsizing across the full
range of $M_*\sim 10^{9.5-12} M_\odot$, namely that more massive
galaxies contain stars that on average formed earlier.  The most
massive galaxies have a median stellar age of $\approx 9-10$~Gyr
($z\sim 2$), while small dwarfs show a more protracted, constant
or increasing star formation history with a median stellar age of
$\approx 5-6$~Gyr ($z\sim 0.6$).

\item The predicted star formation rate--stellar mass relation (main
sequence) is in general agreement with recent H$\alpha$ and UV-based
SFR data at $z\approx 0$.  But as with previous models, \mufasa\
has difficulty matching observations at $z\sim 1-3$, falling short
in specific star formation rates at $z=2$ by $\sim\times 3$.  It
remains a perplexing mystery how the overall GSMF evolution can be
so well reproduced, while a direct measure of growth rates in
individual galaxies suggests a significant discrepancy through the
peak epoch of galaxy growth.

\item Comparing GSMFs between identical runs using MFM, P-SPH, and
T-SPH shows that hydrodynamics methodology plays a small but
non-negligible role in determining the GSMF.  It is increasingly
influential at late times and in more massive galaxies, presumably
because this is where the interactions between outflows and accreting
gas become more important for galaxy growth.  Nonetheless, such
variations are sub-dominant compared to variations in the GSMF owing
to choices in feedback.  For instance, our two-phase winds including
a 30\% supernova-heated component adds substantial preventive
feedback that suppresses the GSMF by $\sim\times 2-3$ by today.

\end{itemize}

The \mufasa\ simulations represent a continued shift in the approach
towards incorporating subgrid physics into cosmological-scale
simulations of galaxy formation.  \mufasa\ relies increasingly on
the results from very high-resolution zoom simulations and
well-constrained analytic models to constrain subgrid prescriptions,
rather than simple but highly tuned parameterizations.  In this way
it takes advantage of the true multi-scale nature of simulations
today, marrying models spanning far greater dynamic range than can
be represented in any single simulation.  \mufasa's success at
reproducing key galaxy observables over most of cosmic time
demonstrates that this is a fruitful approach.

Despite being substantially less fine-tuned than many previous
galaxy formation simulations, we still have free parameters in our
outflow modeling that could be more tightly constrained from external
models, such as the outflow velocity and thermal state.  Furthermore,
our current prescription for quenching massive galaxies remains
purely heuristic, so must be transitioned to a more physically-based
model that includes the self-consistent growth of and energy release
from central black holes.  As our ability to model the physics
driving feedback owing to star formation and active galactic nuclei
grows, the goal is to replace ad hoc subgrid prescriptions with
direct characterisations from more physically-based models.  The
\mufasa\ simulations presented here offer an important step in this
direction, while providing a state-of-the-art platform to investigate
the physical processes driving galaxy evolution across a wide range
of mass scales and cosmic epochs.

 \section*{Acknowledgements}
The authors thank D. Angl\'es-Alc\'azar, F. Durier, S. Huang, N.
Katz, T. Naab, B. Oppenheimer, M. Rafieferantsoa, and J. Schaye for
helpful conversations.  The authors thank Paul Torrey for providing
us with the Illustris data, Rob Crain and Joop Schaye for the EAGLE
data, and Ananth Tenneti and Tiziani DiMatteo for the MassiveBlack-II
data.  RD and RJT acknowledge support from the South African Research
Chairs Initiative and the South African National Research Foundation.
Support for RD was also provided by NASA ATP grant NNX12AH86G to
the University of Arizona.  Supported for RJT was provided in part
by the Gordon and Betty Moore Foundation’s Data-Driven Discovery
Initiative through Grant GBMF4561 to Matthew Turk, and by the
National Science Foundation under grant \#ACI-1535651.  Support
for PFH was provided by an Alfred P. Sloan Research Fellowship,
NASA ATP Grant NNX14AH35G, and NSF Collaborative Research Grant
\#1411920 and CAREER grant \#1455342.  The simulations were run on
the Pumbaa astrophysics computing cluster hosted at the University
of the Western Cape, which was generously funded by UWC's Office
of the Deputy Vice Chancellor.  These \mufasa\ simulations were run
with revision e77f814 of {\sc Gizmo} hosted at {\tt
https://bitbucket.org/rthompson/gizmo}.


\begin{thebibliography}{1}
\bibitem[Agertz et al.(2007)]{Agertz-07} Agertz, O. et al. 2007, MNRAS, 380, 963
\bibitem[Angl\'es-Alc\'azar et al.(2015)]{Angles-15} Angl\'es-Alc\'azar, D., \"Ozel, F., Dav\'e, R., Katz, N., Kollmeier, J. A., Oppenheimer, B. D. 2015, 800, 127
\bibitem[Angl\'es-Alc\'azar et al.(2016)]{Angles-16} Angl\'es-Alc\'azar, D., Dav\'e, R., Faucher-Giguere, C.A., \"Ozel, F., Hopkins, P.F. 2016, MNRAS, submitted, arXiv:1603.08007 
\bibitem[Asplund et al.(2009)]{Asplund-09} Asplund, M., Grevesse, N., Sauval, A. J., Scott, P. 2009, ARA\&A, 47, 481
\bibitem[Baldry et al.(2012)]{Baldry-12} Baldry, I. K. et al. 2012, MNRAS, 421, 621
\bibitem[Bastian, Covey, \& Meyer(2010)]{Bastian-10} Bastian, N., Covey, K. R., Meyer, M. R. 2010, ARA\&A, 48, 339
\bibitem[Bauer et al.(2013)]{Bauer-13} Bauer, A.E., Hopkins, A.M., Gunawardhana, M. et al. 2013, MNRAS, 434, 209
\bibitem[Behroozi et al.(2013)]{Behroozi-13} Behroozi, P. S., Wechsler, R. H., Conroy, C. 2013, ApJ, 770, 57
\bibitem[Bell et al.(2003)]{Bell-03} Bell, E. F., McIntosh, D. H., Katz, N., Weinberg, M. D. 2003, ApJS, 149, 289
\bibitem[Bruzual \& Charlot(2003)]{Bruzual-03} Bruzual, G., Charlot, S. 2003, MNRAS, 344, 1000
\bibitem[Caputi et al.(2011)]{Caputi-11} Caputi, K. I., Cirasuolo, M., Dunlop, J. S., McLure, R. J., Farrah, D., Almaini, O. 2011, MNRAS, 413, 162
\bibitem[Chabrier(2003)]{Chabrier-03} Chabrier G., 2003, PASP, 115, 763
\bibitem[Chang et al.(2015)]{Chang-15} Chang, Y.-Y., van der Wel, A., da Cunha, E., Rix, H.-W. 2015, ApJS, 219, 8
\bibitem[Christensen et al.(2016)]{Christensen-16} Christensen, C. Dav\'e, R., Governato, F., Pontzen, A., Brooks, A., Munshi, F., Quinn, T., Wadsley, J. 2015, MNRAS, submitted, arXiv:1508.00007
\bibitem[Conroy, van Dokkum \& Kravtsov(2015)]{Conroy-15} Conroy, C., van Dokkum, P.G., Kravtsov, A. 2015, ApJ 803, 77
\bibitem[Cowie et al.(1996)]{Cowie-96} Cowie, L. L., Songaila, A., Hu, E. M., Cohen, J. G. 1996, AJ, 112, 839
\bibitem[Crain et al.(2015)]{Crain-15} Crain, R. A. 2015, MNRAS, 450, 1937
\bibitem[\protect\citeauthoryear{Croton et al.} {2006}]{Croton-06} Croton, D.~J., et al. 2006, MNRAS, 365, 11
\bibitem[Cullen \& Dehnen(2010)]{Cullen-10} Cullen, L., Dehnen, W. 2010, MNRAS, 408, 669
\bibitem[Daddi et al.(2007)]{Daddi-07} Daddi, E., et al. 2007, ApJ, 670, 156
\bibitem[Dav\'e(2008)]{Dave-08} Dav\'e, R. 2008, MNRAS, 385, 147
\bibitem[Dav\'e, Oppenheimer, \& Finlator(2011)]{Dave-11a} Dav\'e, R., Oppenheimer, B. D., Finlator, K. M. 2011, MNRAS, 415, 11
\bibitem[Dav\'e, Finlator, \& Oppenheimer(2011)]{Dave-11b} Dav\'e, R., Finlator, K. M., Oppenheimer, B. D. 2011, MNRAS, 416, 1354
\bibitem[Dav\'e et al.(2013)]{Dave-13} Dav\'e, R., Katz, N., Oppenheimer, B. D., Kollmeier, J. A., Weinberg, D. H. 2013, MNRAS, 434, 2645
\bibitem[Dehnen \& Aly(2012)]{Dehnen-12} Dehnen, W., Aly, H. 2012, MNRAS, 425, 1068
\bibitem[Dekel \& Silk(1986)]{Dekel-86} Dekel, A. \& Silk, J. 1986, ApJ, 303, 39
\bibitem[Duncan et al.(2014)]{Duncan-14} Duncan, K. 2014, MNRAS, 444, 2960
\bibitem[Durier \& Dalla Vecchia(2012)]{Durier-12} Durier, F., Dalla Vecchia, C. 2012, MNRAS, 419, 465
\bibitem[The Enzo Collaboration(2014)]{Enzo-14} The Enzo Collaboration, Bryan, G. L., Norman, M. L., et al. 2014, ApJS, 211, 19
\bibitem[Duffy et al.(2008)]{Duffy-08} 	Duffy, A.R., Schaye, J., Kay, S.T., Dalla Vecchia, C. 2008, MNRAS, 390, L64
\bibitem[Faucher-Giguere, Kere\v{s}, \& Ma(2009)]{Faucher-09} Faucher-Giguere, C. A., Lidz, A., Zaldarriaga, M., Hernquist, L. 2009, ApJ, 703, 1416
\bibitem[Faucher-Giguere, Kere\v{s}, \& Ma(2010)]{Faucher-10} Faucher-Giguere, C. A., Kere\v{s}, D., Dijkstra, M., Hernquist, L., Zaldarriaga, M. 2010, ApJ, 725, 633
\bibitem[Ferland (2004)]{Ferland-04} Ferland, G. J. 2004, Bull. of the AAS, 36, 1574
\bibitem[Ford et al. (2016)]{Ford-16} Ford, A.B., Werk, J.K., Dav\'e, R. et al. 2016, MNRAS, accepted, arXiv:1503.02084
\bibitem[Gabor \& Dav\'e(2010)]{Gabor-10} Gabor, J. M., Dav\'e, R. 2010, MNRAS, 407, 749
\bibitem[Gabor et al.(2011)]{Gabor-11} Gabor, J. M., Dav\'e, R., Oppenheimer, B. D., Finlator, K. M. 2011, MNRAS, 417, 2676
\bibitem[Gabor \& Dav\'e(2012)]{Gabor-12} Gabor, J. M. \& Dav\'e, R. 2012, MNRAS, 427, 1816
\bibitem[Gabor \& Dav\'e(2015)]{Gabor-15} Gabor, J. M. \& Dav\'e, R. 2015, MNRAS, 447, 374
\bibitem[Genel et al.(2014)]{Genel-14} Genel, S. et al. 2014, MNRAS, 445, 175
\bibitem[Gonzalez et al.(2011)]{Gonzalez-11} Gonz\'alez, V., Labb\'e, I., Bouwens, R.J., et al. 2011, ApJL, 735, L34
\bibitem[Governato et al.(2007)]{Governato-07} Governato, F., Willman, B., Mayer, L. 2007, MNRAS, 374, 1479
\bibitem[Grazian et al.(2015)]{Grazian-15} Grazian, A. et al. 2015, A\&A, 575, 96
\bibitem[Grogin et al.(2011)]{Grogin-11} Grogin, N. A. et al. 2011, ApJS, 197, 35
\bibitem[Haardt \& Madau(2012)]{Haardt-12} Haardt, F., Madau, P. 2012, ApJ, 746, 125
\bibitem[Hahn \& Abel(2011)]{Hahn-11} Hahn, O., Abel, T. 2011, MNRAS, 415, 2101
\bibitem[Henriques et al.(2015)]{Henriques-15} Henriques, B. M. B., White, S. D. M., Thomas, P. A., Angulo, R., Guo, Q., Lemson, G., Springel, V., Overzier, R. 2015, MNRAS, 451, 2633
\bibitem[Hernquist \& Katz(1989)]{Hernquist-89} Hernquist, L. \& Katz, N. 1989, ApJS, 70, 419
\bibitem[Hopkins(2013)]{Hopkins-13} Hopkins, P.F. 2013, MNRAS, 428, 2840
\bibitem[Hopkins et al.(2014)]{Hopkins-14} Hopkins, P.F., Kere\v{s}, D., O\~norbe, J., Faucher-Giguere, C. A.,Quataert, E., Murray, N., Bullock, J. S. 2014, MNRAS, 445, 581
\bibitem[Hopkins(2015)]{Hopkins-15a} Hopkins, P.F. 2015, MNRAS, 450, 53
\bibitem[Hopkins \& Raives(2015)]{Hopkins-15b} Hopkins, P.F., Raives, M.J. 2015, arXiv:1505.02783
\bibitem[Ilbert et al.(2013)]{Ilbert-13} Ilbert, O. et al. 2013, A\&A, 556, 55
\bibitem[Iwamoto et al.(1999)]{Iwamoto-99} Iwamoto, K., Brachwitz, F., Nomoto, K., Kishimoto, N., Umeda, H., Hix, W. R., Thielemann, F.-K. 1999, ApJS, 125, 439
\bibitem[Kennicutt(1998)]{Kennicutt-98} Kennicutt, R. C. 1998, ApJ, 498, 541
\bibitem[Kere\v{s} et al.(2005)]{Keres-05} Kere\v{s}, D., Katz, N., Weinberg, D. H., \& Dav\'e, R. 2005, MNRAS, 363, 2
\bibitem[Khandai et al.(2015)]{Khandai-15} Khandai, N., Di Matteo, T., Croft, R., Wilkins, S., Feng, Y., Tucker, E., DeGraf, C., Liu, M.-S. 2015, MNRAS, 450, 1349
\bibitem[Kim et al.(2014)]{Kim-14} Kim, J.-h., Abel, T., Agertz, O., et al. 2014, ApJS, 210, 14
\bibitem[Koekemoer et al.(2011)]{Koekemoer-11} Koekemoer, A. M. et al. 2011, ApJS, 197, 36
\bibitem[Kollmeier et al.(2014)]{Kollmeier-14} Kollmeier, J. A. et al. 2014, ApJ, 789, L32
\bibitem[Kravtsov, Vikhlinin, \& Meshscheryakov(2014)]{Kravtsov-14} Kravtsov, A., Vikhlinin, A., Meshscheryakov, A. 2014, ApJ, submitted, arXiv:1401.7329
\bibitem[Kriek et al.(2009)]{Kriek-09} Kriek, M., van Dokkum, P. G., Franx, M., Illingworth, G. D., Magee, D. K. 2009, ApJ, 705, 71
\bibitem[Krumholz, McKee, \& Tumlinson(2009)]{Krumholz-09} Krumholz, M. R., McKee, C. F., Tumlinson, J. T. 2008, ApJ, 693, 216
\bibitem[Krumholz \& Gnedin(2011)]{Krumholz-11} Krumholz, M. R., McKee, C. F., Tumlinson, J. T. 2011, ApJ, 729, 36
\bibitem[\L okas et al.(2001)]{Lokas-01} \L okas, E.L., Mamon, G.A. 2001, MNRAS, 321, 155
\bibitem[Lee et al.(2012)]{Lee-12} Lee, K.-S. et al. 2012, ApJ, 752, 66
\bibitem[Li \& White(2009)]{Li-09} Li, C., White, S.D.M. 2009, MNRAS, 398, 2177
\bibitem[Liang et al.(2016)]{Liang-16} Liang, L., Durier, F., Babul, A., Dav\'e, R., Oppenheimer, B. D., Katz, N., Fardal, M., Quinn, T. 2016, MNRAS, 456, 4266
\bibitem[Madau \& Dickinson(2014)]{Madau-14} Madau, P., Dickinson, M. 2014, ARA\&A, 52, 415
\bibitem[Martin(2005)]{Martin-05} Martin, C. L. 2005, ApJ, 621, 227
\bibitem[McAlpine et al.(2015)]{McAlpine-15} McAlpine, S., Helly, J.C., Schaller, M. et al. 2015, arXiv:1510.01320
\bibitem[McGaugh et al.(2012)]{McGaugh-12} McGaugh, S.S. 2012, AJ, 143, 40
\bibitem[Mitra, Dav\'e, \& Finlator(2015)]{Mitra-15} Mitra, S., Dav\'e, R., Finlator, K. 2015, MNRAS, 452, 1184
\bibitem[Mo, Mao, \& White(1998)]{Mo-98} Mo, H.J., Mao, S., White, S.D.M. 1998, MNRAS, 295, 319
\bibitem[Mobasher et al.(2015)]{Mobasher-15} Mobasher, B. 2015, ApJ, 808, 101
\bibitem[Monaghan(1992)]{Monaghan-92} Monaghan, J. J. 1992, ARA\&A, 30, 543
\bibitem[Moustakas et al.(2014)]{Moustakas-13} Moustakas, J. et al. 2013, ApJ, 767, 50
\bibitem[Muratov et al.(2015)]{Muratov-15} Muratov, A.L., Kere\v{s}, D.,  Faucher-Giguere, C. A., Hopkins, P. F., Quataert, E., Murray, N. 2015, MNRAS, submitted, arXiv:1501.03155
\bibitem[Muzzin et al.(2013)]{Muzzin-13} Muzzin, A. et al. 2013, ApJ, 777, 18
\bibitem[Navarro, Frenk, \& White(1997)]{Navarro-97} Navarro, J., Frenk, C.S., White, S.D.M. 1997, ApJ, 490, 493
\bibitem[Neistein, Dekel, \& van den Bosch(2006)]{Neistein-06} Neistein, E., van den Bosch, F. C., Dekel, A. 2006, MNRAS, 372, 933
\bibitem[Noeske et al. (2007)]{Noeske-07} Noeske, K. G., et al. 2007a, ApJL, 660, L43
\bibitem[Nomoto et al.(2006)]{Nomoto-06} Nomoto, K., Tominaga, N., Umeda, H., Kobayashi, C., Maeda, K. 2006, NuPhA, 777, 424
\bibitem[Oppenheimer \& Dav\'e(2006)]{Oppenheimer-06} Oppenheimer, B. D. \& Dav\'e, R. 2006, MNRAS, 373, 1265
\bibitem[Oppenheimer \& Dav\'e(2008)]{Oppenheimer-08} Oppenheimer, B. D. \& Dav\'e, R. 2008, MNRAS, 387, 577
\bibitem[Oppenheimer et al.(2010)]{Oppenheimer-10} Oppenheimer, B.D., Dav\'e, R., Kere\v{s}, D., Katz, N., Kollmeier, J.A., Weinberg, D.H. 2010, MNRAS, 406, 2325
\bibitem[Planck(2015)]{Planck-15} Planck Collaboration, Ade, P. A. R., Aghanim, N., et al. 2015, arXiv:1502.01589
\bibitem[Popping et al.(2009)]{Popping-09} Popping, A., Dav\'e, R., Braun, R., Oppenheimer, B. D. 2009, A\&A, 504, 15
\bibitem[Rahmati et al.(2013)]{Rahmati-13} Rahmati, A., Pawlik, A. H., Rai\v{c}evi\'c, M., Schaye, J.
\bibitem[Read \& Hayfield(2012)]{Read-12} Read, J. I., Hayfield, T. 2012, MNRAS, 422, 3037
\bibitem[Reddy \& Steidel(2009)]{Reddy-09} Reddy N.A., Steidel, C.C. 2009, ApJ, 692, 778
\bibitem[Reddy et al.(2012)]{Reddy-12} Reddy N.A., et al. 2012. ApJ 744, 154
\bibitem[Rodighiero et al.(2011)]{Rodighiero-11} Rodighiero, G., Daddi, E., Baronchelli, I. 2011, ApJ, 739, L40
\bibitem[Salim et al. (2007)]{Salim-07} Salim, S. et al. 2007, ApJS, 207, 173, 267
\bibitem[Salmon et al. (2015)]{Salmon-15} Salmon, B., Papovich, C., Finkelstein, S.L. et al. 2015, ApJ, 799, 183
\bibitem[Scannapieco \& Bildsten(2005)]{Scannapieco-05} Scannapieco, E. \& Bildsten, L. 2005 ApJ, 629, L85
\bibitem[Schaller et al.(2015)]{Schaller-15} Schaller, M., Dalla Vecchia, C., Schaye, J., Bower, R. G., Theuns, T., Crain, R. A., Furlong, M., McCarthy, I. G. 2015, MNRAS, 454, 2277
\bibitem[Schaye, Carswell, \& Kim(2007)]{Schaye-07} Schaye, J., Carswell, R.F., Kim, T.-S. 2007, MNRAS, 379, 1169
\bibitem[Schaye \& Dalla Vecchia(2008)]{Schaye-08} Schaye, J. \& Dalla Vecchia, C. 2008, MNRAS, 383, 1210
\bibitem[Schaye et al.(2010)]{Schaye-10} Schaye, J., Dalla Vecchia, C., Booth, C.M. 2010, MNRAS, 402, 1536
\bibitem[Schaye et al.(2015)]{Schaye-15} Schaye, J. et al. 2015, MNRAS, 446, 521
\bibitem[\protect\citeauthoryear{Schmidt} {1959}]{Schmidt-59} Schmidt, M.\ 1959, ApJ, 129, 243
\bibitem[Sijacki et al.(2007)]{Sijacki-07} Sijacki, D. Springel, V., Di Matteo, T., Hernquist, L. 2007, MNRAS, 380, 877
\bibitem[Somerville et al.(2008)]{Somerville-08} Somerville, R. S., Hopkins, P. F., Cox, T. J., Robertson, B. E., Hernquist, L. 2008, MNRAS, 391, 481
\bibitem[Somerville \& Dav\'e(2015)]{Somerville-15} Somerville, R. S., Dav\'e, R. 2015, ARA\&A, 53, 51
\bibitem[Song et al.(2015)]{Song-15} Song, M. et al. 2015, ApJ, 791, 3
\bibitem[Sparre et al.(2015)]{Sparre-15} Sparre, M. et al. 2015, MNRAS, 447, 3548
\bibitem[Speagle et al.(2014)]{Speagle-14} Speagle, J. S., Steinhardt, C. L., Capak, P. L., Silverman, J. D. 2014, ApJS, 214, 15
\bibitem[Springel \& Hernquist(2002)]{Springel-02} Springel, V. \& Hernquist, L. 2002, MNRAS, 333, 649
\bibitem[Springel \& Hernquist(2003)]{Springel-03} Springel, V. \& Hernquist, L. 2003, MNRAS, 339, 289
\bibitem[Springel \& Hernquist(2003)]{Springel-03b} Springel, V. \& Hernquist, L. 2003, MNRAS, 339, 312
\bibitem[Springel(2005)]{Springel-05} Springel, V. 2005, MNRAS, 364, 1105
\bibitem[Springel(2010)]{Springel-10} Springel, V. 2010, MNRAS, 401, 791
\bibitem[Stark et al.(2013)]{Stark-13} 	Stark, D. P., Schenker, M. A., Ellis, R., Robertson, B., McLure, R., Dunlop, J. 2013, ApJ, 763, 129
\bibitem[Steidel et al.(2010)]{Steidel-10} Steidel, C. C. et al. 2010, ApJ, 717, 298
\bibitem[Stinson et al.(2013)]{Stinson-13} Stinson, G. S., Brook, C., Maccio, A. V., Wadsley, J., Quinn, T. R., Couchman, H. M. P. 2013, MNRAS, 428, 129
\bibitem[Sullivan et al.(2006)]{Sullivan-06} Sullivan, M. et al. 20016, ApJ, 648, 868
\bibitem[Thomas et al.(2005)]{Thomas-05} Thomas, D., Maraston, C., Bender, R., Mendes de Oliveira, C. 2005, ApJ, 621, 673
\bibitem[Thompson et al.(2014)]{Thompson-14} Thompson, R., Nagamine, K., Jaacks, J., Choi, J.-H. 2014, ApJ, 780, 145
\bibitem[Thompson(2015)]{Thompson-15} Thompson, R. 2015, ASCL, 1502, 12
\bibitem[Thompson et al.(2016)]{Thompson-16} Thompson, R., Dav\'e, R., Huang, S., Katz, N., MNRAS, submitted, arXiv:1508.01851
\bibitem[Tomczak et al.(2014)]{Tomczak-14} Tomczak, A. R. 2014, ApJ, 783, 85
\bibitem[Tumlinson et al.(2013)]{Tumlinson-13} Tumlinson, J., Thom, C., Werk, J.K. et al. 2013, ApJ, 777, 59
\bibitem[Vogelsberger et al.(2014)]{Vogelsberger-14} Vogelsberger, M., Genel, S., Springel, V., Torrey, P., Sijacki, D., Xu, D., Snyder, G., Nelson, D., Hernquist, L. 2014, MNRAS, 444, 1518
\bibitem[Voit(2005)]{Voit-05} Voit, G.M. 2005, AdSpR, 36, 701
\bibitem[Voit et al.(2015)]{Voit-15} Voit, G. M., Donahue, M., Bryan, G. L., McDonald, M. 2015, Nature, 519, 203
\bibitem[Voit \& Donahue(2015)]{Voit-15b} Voit, G. M., Donahue, M. 2015, ApJL, 799, L1
\bibitem[Wetzel et al.(2016)]{Wetzel-16} Wetzel, A. R., Hopkins, P. F., Kim, J.-h., Faucher-Giguere, C.-A., Keres, D., Quataert, E. 2016, ApJL, submitted, arXiv:1602.05957
\bibitem[White et al.(2015)]{White-15} White, C. E., Somerville, R. S., Ferguson, H. C. 2015, ApJ, 799, 201
\bibitem[Whitaker et al.(2014)]{Whitaker-14} Whitaker, K.E. et al. 2014, ApJ, 795, 104
\bibitem[Wilkins, Trentham, \& Hopkins(2008)]{Wilkins-08} Wilkins, S. M., Trentham, N., Hopkins, A. M. 2008, MNRAS, 385, 687
\bibitem[Wilkinson et al.(2016)]{Wilkinson-16} Wilkinson, A., Almaini, O., Chen, C.-C. et al. 2016, MNRAS, submitted, arXiv:1604.00018

\end{thebibliography}
\end{document}